\documentclass[twocolappendix,floatfix]{emulateapj}
\usepackage{amsmath,amssymb, graphicx, setspace}
\usepackage{hyperref}
\usepackage{physics}
\usepackage{commath}
\usepackage{float}
\usepackage{subfigure}
\usepackage{eqnarray,amsmath}
\usepackage{xcolor,cancel}
\usepackage[normalem]{ulem}
\usepackage{url}
\usepackage{natbib}

\usepackage{lineno}
\usepackage{cleveref}
\crefname{section}{Section}{Sections}
\crefname{figure}{Figure}{Figures}
\crefname{table}{Table}{Tables}
\crefname{equation}{Eq.}{Eq.}

\begin{document}
\shortauthors{Shah et. al.}
\shorttitle{Frequency plan for WDBs}
\title{On the optimal Frequency plan for LISA pre-science operations using verification binaries}
\author{Sweta Shah$^{1}$ Valeriya Korol$^{2}$, Thomas Kupfer$^{3,4}$}
\affil{$^1$Leibniz Universit\"at Hannover, Institut f\"ur Gravitationsphysik, Callinstra{\ss}e 38, Hannover 30167, Germany\\
 $^2$Max-Planck-Institut f\"ur Astrosphysik, Karl-Schwarzschild-Stra{\ss}e 1, Garching 85748, Germany\\
 $^3$Hamburger Sternwarte, University of Hamburg, Gojenbergsweg 112, D-21029 Hamburg, Germany\\
 $^4$Department of Physics and Astronomy, Texas Tech University, P.O. Box 41051, Lubbock, TX 79409, USA}
\email{sweta.shah@aei.mpg.de}
\begin{abstract}
The future Laser Interferometer Space Antenna (LISA) mission, which has successfully passed Mission Formulation phase, is in planning to be launched in 2030s. One of the ubiquitous LISA sources are the white-dwarf binaries (WDB) of which $\sim$40 are guaranteed sources as of now, making LISA unique in comparison to its ground-based counterpart. The current hardware design in planning necessitates a thorough check to determine whether the various locking schemes influence the guaranteed sources' signals significantly. This could have implication to re-consider the choice of the initial locking configuration that is optimal for the pre-science or science operations phase. Comparison of the phasemeter output of a face-on (V407Vul) binary and an edge-on (ZTFJ2243 binary indicates that the non-swap locking scheme with the maximal Doppler locks, is optimal for instrument calibration. The most well known locking scheme from Shaddock et al (2004) gives an TDI Michelson SNR for VB that is larger by factor of 10 in comparison to free-running laser configuration. Comparison with another locking scheme with most number of doppler links in the phasemeter measurement gives SNR difference of $\sim 30\%$ and $\sim 20\%$ for the face-on system V407Vul and the edge-on system ZTFJ2243 respectively. We find similar amplitudes in the TDI output stream for the face-on system V407Vul and the edge-on system ZTFJ2243 which leads to a significantly smaller inclination bias for the non-swap locking scheme. Additionally, a larger amplitude for edge-on systems will benefit most verification systems as the population of verification systems is biased towards edge-on systems as they are easier to detect in electromagnetic data.
\end{abstract}
\maketitle
\section{Introduction} \label{intro}
Gravitational wave (GW) astronomy has become a norm since its first direct detection in 2015, announced in 2016 \citep{bbhPRL2016}. This provided a strong push forward for the long envisaged space-based GW detector concept, the Laser Interferometer Space Antenna, LISA \footnote{\url{https://www.elisascience.org/}}, an European Space Agency (ESA) \textit{flagship} mission with a contribution from the National Aeronautics and Space Administration (NASA) and with a planned launch year in the late $\sim$2030s.

With its well-defined science cases \citep{2023LRR}, the detector's sensing window for 4-dimensional space-time fluctuations has been set to a range of $[10^{-4}-1]$ Hertz (Hz). This characteristic band not only complements its ground-based counterparts, but also allows LISA to cover a broader science case encompassing larger ranges of source masses, mass-ratios, and redshifts in astrophysics and cosmography. The most numerous astronomical class of sources observable by LISA are the compact White Dwarf Binaries (WDBs) in our home-galaxy, the \textit{Milky-Way}. Population synthesis models predict the detector to be able to individually resolve ${\cal O}(10^4)$ sources \citep{Kupfer2018}, detached and mass-transferring \citep[e.g.][]{Toonen2014}, from an unresolvable foreground comprising millions of compact Galactic binaries which, at $\sim 3\times10^{-3}$ Hz overtakes the noise from the instrument \citep[e.g.][]{Nissanke2012}.

The ground-breaking key technology of needing the test mass in free-fall at the level of an unfathomable $\sim 15\times 10^ {-15}$m$s^{-2}/\sqrt{\text{Hz}}$\footnote{$10^ {-15}$ is 1 femtometer} to achieve LISA's scientific goals for the frequency band stated above has been demonstrated successfully by LISA Pathfinder, LISA's precursor mission by ESA \citep{femtoPRL2016}. Another crucial key technology for long-arm interferometry has been partially demonstrated with the Gravity Recovery And Climate Experiment Follow-On (GRACE-FO). This Earth observation mission employed twin satellites and a 220km Laser Ranging Interferometer instrument \citep{GraceFO2019} with a laser whose wavelength is 1064 nano-meters \citep{Koch2020}, which is planned to be used for LISA as well.

The practicality of the measurement of real numbers in the phasemeters of LISA \citep{Barke2015} requires careful planning of the laser frequencies that the constellation will carry onboard its three spacecraft (S/C), where each spacecraft houses two optical benches (OB) constructed from low-expansion ceramic material\citep{Troebs2019} whereupon the interferometric measurements will occur. The reason for frequency plan stems from the fundamental need for heterodyne interferometry between two laser beams, where one of them has a power level in the order of sub nano-Watt with respect to the other with unequal frequencies. These constellation orbits \citep{Martens2021} introduce a strong dependence on time-varying length changes, which are caused by the imbalance of gravitational pulls from major planets and the Sun. Additionally, the long-lived WDB signals will contribute small variations in amplitude, frequency, and phase.
\begin{figure*}
    \begin{minipage}[h!]{\textwidth}
        \includegraphics[width=\textwidth]{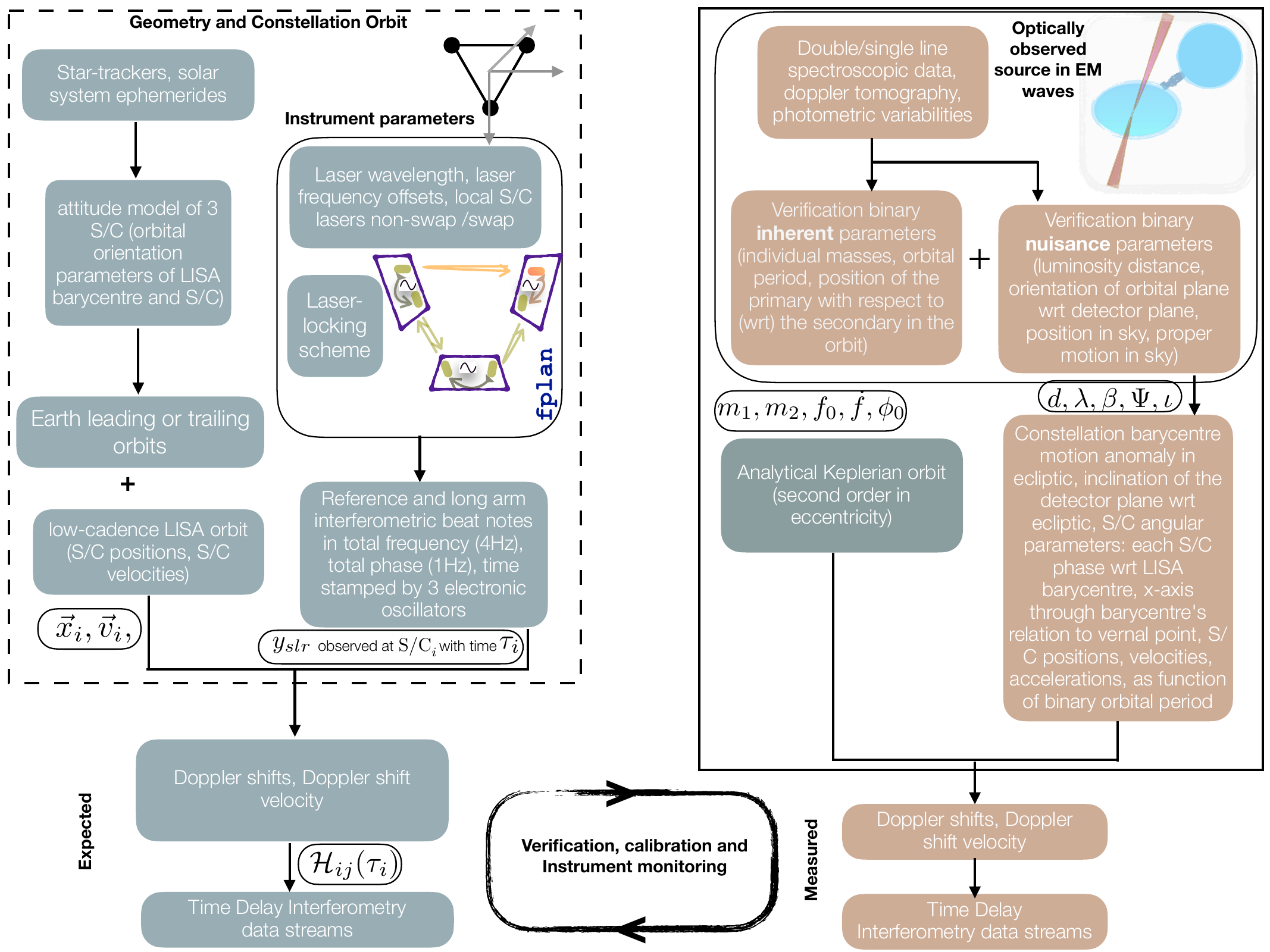}
        \caption{LISA constellation geometry from star trackers and S/C constituent properties verified against those observed from long-lived, stable Verification (\textit{apriori known} white dwarf) binaries. Left part of the diagram gives the instrument model with the parameters describing them whereas right part of the diagram gives the flow of the optically measured VB binary parameters implicating for the former.}
        \label{fig:main_algo}
    \end{minipage}
    \hfill
\end{figure*}
There are fundamentally two sources of constraints on the design of the allowed frequencies of the laser. One is the noise from the transimpedance amplifiers in the photodiode. This noise limits the photodiode's capability to read out the phases of the heterodyne interferometric signal (with unequal frequencies) above 25MHz, leading to increased noise \citep{Barranco2018}. Below 5MHz, the sensors are inundated with relative intensity noise (RIN) from the laser. RIN couples into the interferometric readout substantially at the heterodyne frequency violating the required noise level \citep{Wissel2022b}, which is mitigated in hardware by a technique, the so-called \textit{balanced-detection} \citep{Wissel2022a}. 

Secondly, the phasemeter cannot distinguish negative (blue-shifted) or positive (red-shifted) Doppler shifts from the sinusoidal changes\footnote{also known as  polarity of a laser frequency offset in the Instrument \textit{lingua-franca}} of the armlengths between any two S/C nor the minute Doppler shifts from the WDBs and is not capable of dealing with zero crossing values where the two interfered lasers to have identical frequencies. The space-qualified phasemeter designed for LISA \cite{Gerberding2013} has therefore strict constraints from photodiodes' sensitivity and phasemeter's internal limitations.

From these hardware restraints, three distinct ways to form locks between six independent lasers on-board have been discovered earlier by \citep{Shaddock2004},~\citep{Barke2015} for the setup where any two lasers on the two OBs within one S/C are not allowed to interfere with the received laser from the distant S/C in forming the 2.5 million km armlength heterodyne signal. \textit{Locking} essentially means imprinting a copy of an interfered signal's phase (or frequency) to that of the local laser beam with a pre-determined offset comparable to that of the long-arm Doppler shift. The three solutions proliferate for the generic set up of the lasers in how they can be locked to each other when all allowable frequencies by the nominal hardware design are imposed in a brute-force exploration of interferometric signals' frequency parameter space \citep{Heinzel2020}, known as \textit{frequency plan} in LISA, commonly referred by, ``\texttt{fplan}".

A known GW signal's will provide an independent calibration of the instrument sub-systems, in addition to the in-built mechanisms in the hardware. Furthermore they will provide an estimate for the S/C angular parameters distinct than the on board star trackers and on-ground ranging amongst others. The incoherence in the beatnote signals in the form of glitch due to the instrument sub-systems can be caused by laser cycle slip \citep{Gerberding2013} and/or frequency jump in the electronic oscillator to time-stamp the optical signals \citep{Yamamoto2022}. The motivation for using such a signal's inference for the instrument is crudely sketched in \cref{fig:main_algo} with an exemplary case of \texttt{fplan} underlying the topology of the frequencies in LISA constellation\footnote{inspired by calibration schemes in other space missions, such as \textit{Gaia} \cite[Fig. 1,][]{Lindegren2012}.}. A number of previous studies, \citep[e.g.][]{Littenberg2018} have shown the importance and usefulness of WDBs in calibrating the signal's amplitude and phase due to scheduled gaps and unintentional glitches. It is shown by \cite{Littenberg2018} in a Bayesian data analysis that coherent signals of about ten WDBs can be used to estimate an unknown period of gap within an accuracy of 0.1 to 0.2 seconds from a month to a year of data utilising the near constant frequency GW signals of the WDBs. This \textit{proof-of-concept} study utilised Time Delay Interfetometry (TDI) \citep[e.g.,][]{Armstrong1999}, which is a technique optimizing GW signal over instrument noise for a generic setup of six free-running lasers across the constellation. This needs verification against the different choices of \texttt{fplan}, where the coupling of a GW signal bears \textit{no symmetry} and is radically different in any of the various choices as explained later. 

Additionally, it is not immediately clear which of the numerous laser frequency configurations in \citep{Heinzel2020} is optimal with respect to the instrument characterisation, as motivated and shown in \cref{fig:main_algo}. The instrument can be characterized in two groups of parameters: LISA constellation parameters (orientation) and parameters inherent to the instrument sub-systems, such as laser, clock, Gravitational Reference System, and etc. This paper explores using examples of two distinct laser locking schemes for two of the brightest known WDBs from optical observations, also known as Verification Binaries (VB) \citep{Stroeer2006}, the standard candles for LISA instrument. It could be that the choice of one of the six LISA non-swap \texttt{fplan} should be different for commissioning and calibration phase(s) of the mission versus when the science operations begin. The former would be to maximise instrument knowledge, and later would be to yield maximum scientific output for a certain class of predicted astrophysical sources. For this study, the changes were implemented in Synthetic LISA (\texttt{Synth LISA}) to adapt the code to \texttt{fplan}. \texttt{Synth LISA} was originally developed to study LISA instrument noise as complement to other existing softwares \citep{Vallis2005} to simulate LISA Science process. Our study is complementary to recent developments of instrument simulation with non-Gaussian noise at sub-system level in \cite{Bayle2023} and the contemporary software \texttt{LISANode}\footnote{https://gitlab.in2p3.fr/j2b.bayle/LISANode} relevant for further investigations of the instrument beyond the Adoption phase.

There is perhaps no urgency to have this figured out before the mission's Adoption (planned for 2024), as the meticulous design of the phasemeter \citep{Yamamoto2023} already takes that possibility into account. We mean that all the six unique locking schemes can be telecommanded to reconfigure the measurement scheme aboard in-flight, the choice of which could be driven by optimizing multi-messenger observations or instrument calibration. Consequently, this could also have non-trivial implications for the logistics of the initial leg of the data processing pipeline, the \texttt{I}ntial \texttt{N}oise \texttt{Re}duction \texttt{P}ipeline\footnote{internal to LISA Consortium}, \texttt{INReP}\citep{Wiesner2021}\footnote{planned to be run at the Science Operations Centre (SOC), in the ESA Astronomy Centre, ESAC}, which was successfully demonstrated in the mission's formulation review, essentially passing the LISA mission's study phase.

The paper starts by describing the matter-of-fact known binaries in Sec.~\ref{vbs}, with their responses to LISA laser links described in Sec.~\ref{lisa_resp}. The intricate nature of change of response imposed by \texttt{fplan} is discussed in Sec.~\ref{fplan}, followed by their propagation in a simplified couplets of Michelson-type TDI variables, with conclusion in Sec.~\ref{conclusion}.
\section{LISA Verification Binaries} \label{vbs}
There is a number of guaranteed GW sources in the LISA's frequency band. These are known Galactic binaries with orbital periods $< 1\,$h that have been discovered with electromagnetic observatories \citep[e.g.][]{Kupfer2018}. These sources are typically expected to be detected by LISA within weeks to months following constellation acquisition, as indicated by prior studies \citep[e.g.][]{Shah2012}. The most common of these binaries consist of a combination of a neutron star or white dwarf paired with a compact helium star, white dwarf, or another neutron star. As per the latest data, the known candidates that could potentially be detected by LISA are in the order of several tens of binaries \citep{Kupfer2023}. However, with a considerable number of electromagnetic facilities due to be operational in the decade preceding LISA's launch, the pool of detectable binaries by LISA is projected to exceed 100 \citep{Korol2017}.
\begin{figure}[h!]
  \centering
    \includegraphics[width=0.5\textwidth]{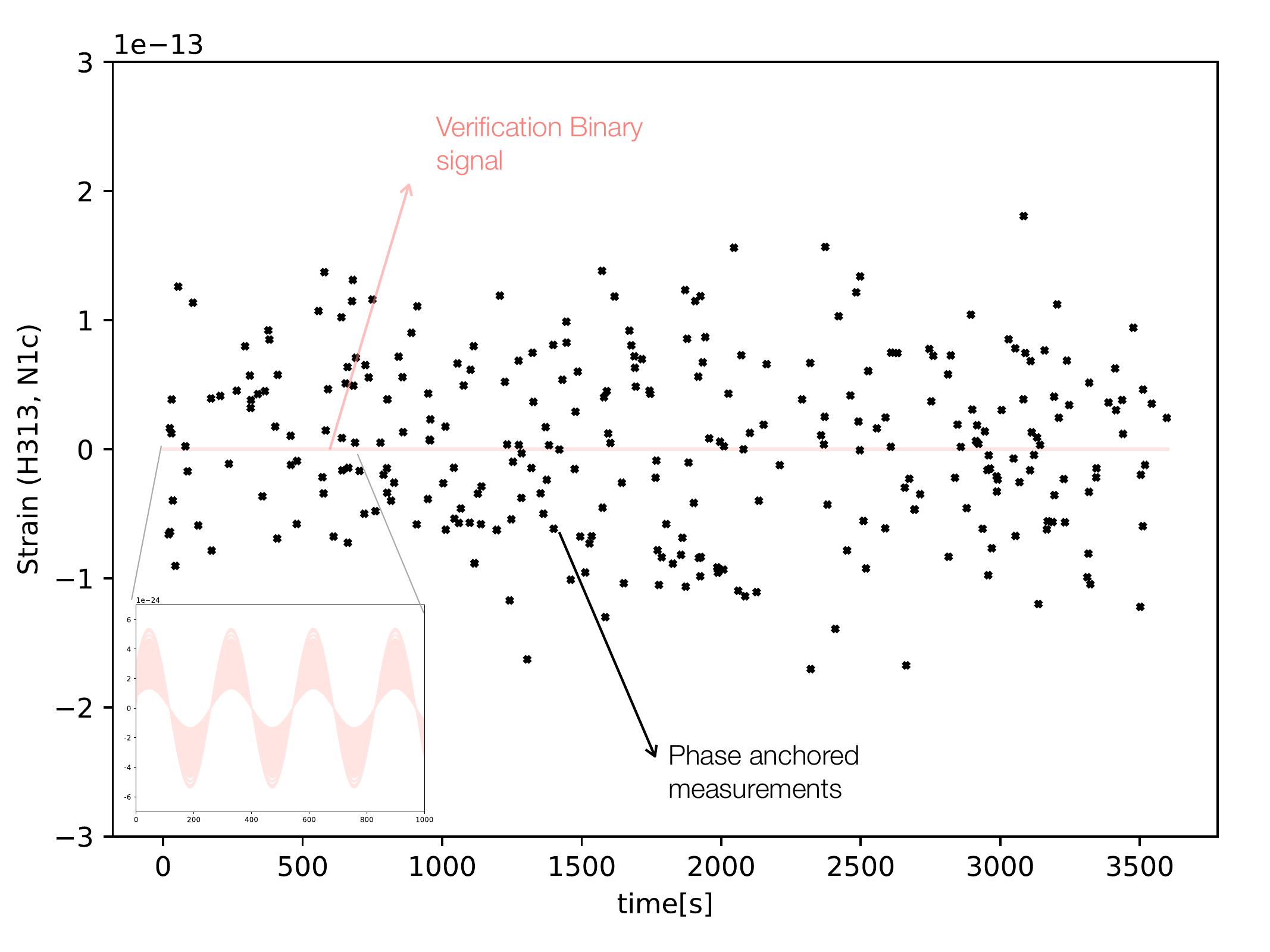}
  \caption{Time series of GW strain from one of the long arm measurements with laser phase-locking for the nominal mission configuration. The predicted signal from the verification binary, V407Vul with errors in its electromagnetic measurements (see Table~\ref{tab:VBsparam}) is shown in rose. Over plotted in black crosses are a simulated example of telemetered data dominated by laser noise where some of the data will be missing potentially due to instrument glitches. The \textit{apriori} known signal from the binary will serve to fill the missing data inherent to instrument for each phasemeter measurement, \textit{independently} to other pre-existing methods. Inset shows a zoom of the predicted signal for 1000s. \label{fig:principle}}
\end{figure}

Known Galactic binaries offer a unique opportunity for verifying the in-orbit performance of LISA, and are thus often referred to as ``verification binaries" (VB) \citep{Stroeer2006}. The primary characteristics of their GW signal, such as amplitude and phase evolution, can be accurately predicted based on their electromagnetic measurements. Consequently, they can be used as tools for monitoring and maintaining the data quality of the LISA instrument. For example, verification binaries with well constrained distances could be used to directly measure the amplitude calibration error and its evolution with time \citep{Savalle2022}.

As mentioned above, the brightest VB can also serve as an independent verification of the instruments' timing standards by demanding coherence across data gaps between data segments \citep{Littenberg2018}. The principle for this and calibration of the instrument data is illustrated in \cref{fig:principle} where the model (to within its estimated errors) for a verification source is shown in red and the inevitable unevenly sampled phasemeter data is shown in black crosses. It's level is dominated by laser frequency noise, expected to be $\sim10\mathcal{O}$ larger than the \textit{apriori} known VB. These measurements need to be combined to form TDI signals to do the analysis in order to suppress the frequency laser noise by $\approx 10$ orders of magnitude \citep{Armstrong1999}. More in general, LISA verification binaries will be especially critical for validating our ability to correctly recover parameters of a source. 

The GW signal from a VB can be described by 8 parameters: 
\begin{equation} \label{eq:VBparam}
    \{\mathcal{A}, f_0, \dot{f}_0, \lambda, \beta, \iota, \psi, \phi_0\}\,, 
\end{equation}
Sky coordinates $(\lambda,\beta)$, the inclination $\iota$, the polarisation $\Psi$ and the amplitude ${\cal A}$ are \textit{extrinsic} parameters that describe the position and orientation of the source with respect to the LISA detector, where \textit{extrinsic} refers to the dependence on detector. The remaining three parameters -- frequency $f$, frequency derivative $\dot{f}$ and initial orbital phase $\phi_0$ -- are \textit{intrinsic}, which implies no dependence on instrument. They determine the temporal evolution of the plus $h_+$ and cross $h_\times$ GW polarisations as follows:
\begin{align} 
    h_+ &= {\cal A}(1+\cos^2\iota) \cos \Phi (t), \nonumber \\
    h_\times &= 2 {\cal A}\cos^2\iota \sin \Phi (t), \\
    \Phi (t) &= 2\pi f_0t +\pi\dot{f}t^2-\phi_0. \nonumber 
\end{align}
The GW amplitude is given by:
\begin{equation} \label{eq:GWamplitude}
    \mathcal{A} = \frac{(G \mathcal{M})^{5/3} }{c^4 d} (\pi f_0)^{2/3}. 
\end{equation}
This is set by the binary's distance, $d$, and chirp mass, given by:
\begin{equation}
    \mathcal{M} = \frac{(m_1 m_2)^{3/5}}{(m_1 + m_2)^{1/5}},  
\end{equation}
for binary component masses $m_1$ and $m_2$. 

Among the eight parameters defining a verification binary's signal (cf. Eq.~\eqref{eq:VBparam}), sky position can be considered to be known with near-perfect precision. This is largely due to electromagnetic observatories' ability to provide micro-arcsecond position measurements. By contrast, LISA's position measurement for verification binaries is estimated to be about a degree \citep[e.g.][]{Finch2023}, a difference of nearly nine orders of magnitude. Recent years have seen substantial progress in determining verification binaries' distances, largely due to astrometric data provided by the \textit{Gaia} mission \citep{gaia2016}; based on a reassessment following the latest data release \citep{gaia2020}, where the current uncertainty ranges between $10^{-3}-10^{-6}$arcsecond \citep{Kupfer2023}.

The orbital period, which relates to GW frequency according to the relationship $f=2/P$, is generally known with high precision, typically to a fractional uncertainty of about $10^{-6}$ (or 0.0001\%), provided that the binary system is oriented edge-on or near edge-on and can be seen as eclipsing. However, for non-eclipsing binaries, the fractional errors can rise to 1\% or less. The measurement of the binary system's inclination is also influenced by its orientation. For eclipsing binaries, the inclination can be determined with sub-degree precision \citep{Burdge2020}. However, for non-eclipsing systems, the inclination remains largely unconstrained. Another parameter that can be determined for eclipsing binary systems is the initial phase of the orbit $\phi_0$, typically assumed at the primary eclipse (the deepest of the two). 
\begin{table}[htpb]
\centering
\begin{tabular}{||c c c c||}
\hline
Params & Units & V407Vul & ZTFJ2243 \\ [0.5ex] 
\hline\hline
$P_\text{orb}$ & s & $569.396230(126)$ &$527.934890(32)$ \\
$\dot{P_\text{orb}}$ & s/s & $-3.12\times 10^{-12}$   & $-2.37\times 10^{-11}$ \\ [1ex] 
$(\lambda, \beta)$ & deg & $(57.7281, 6.4006)$ & $(104.1514, -5.4496)$ \\ [1ex] 
$m_1$    & M$_\odot$ &$0.8\pm0.1$ & $0.349^{+0.093}_{-0.074}$ \\ [1ex] 
$m_2$    & M$_\odot$ &$0.177\pm0.071$ & $0.384^{+0.114}_{-0.074}$ \\ [1ex] 
$d$ & pc & $2089\pm684$ & $1756\pm726$ \\ [1ex] 
$\iota$ & deg & $60$ &  $81.88^{+1.31}_{-0.69}$ \\ [1ex] 
$\mathcal{A}$ & - & $9.38 \times 10^{-23}$ &  $1.22 \times 10^{-22}$ \\ [1ex] 
\hline
\end{tabular}
\caption{Optical photometric, (phase-resolved) spectroscopic measurements derived parameters for the `face-on' V407Vul and the most compact detached know `edge-on' ZTF J1539+5027 binaries \citep{Kupfer2023}. We note that binary's sky coordinates $(\lambda,\beta)$ are given in the heliocentric ecliptic reference frame.}
\label{tab:VBsparam}
\end{table}

Finally, eclipsing systems offer the opportunity to measure the orbital period derivative -- $\dot{P}$ that can be related to GW $\dot{f}$ -- by conducting eclipse timing over an extended time spanning from years to decades depending on the binary's orbital period. This allows for an accurate measurement of the rate of change in the orbital frequency \citep[eg.,][]{Shah2014}. Finally, the challenges in accurately determining the masses of binary components and their distances continue to be the primary sources of uncertainty. Combining these uncertainties, the amplitude ${\cal A}$ in Eq.~\eqref{eq:GWamplitude} can be known to between 1\% and several tens of percent \citep{Kupfer2023}\footnote{For the up-to-date EM parameters, see: \url{https://gitlab.in2p3.fr/LISA/lisa-verification-binaries}}. For this study we select two of the strongest among the currently known verification binaries: one with a face-on orientation with respect to LISA plane, V407~Vul \citep{V407Vul} and the second with an edge-on orientation, ZTF~J2243+5242 \citep[hereafter ZTFJ2243,][]{Burdge2020}. Their measured parameters are 
tabulated in \cref{tab:VBsparam}.
\section{LISA response to Verification Binary GW signal} \label{lisa_resp}
In this section, we focus on how light propagates within the setup of the LISA mission, specifically between the test masses in two inter-S/C, under the influence of weak field gravitational waves. We derive the changes in the frequency of the lasers, exchanged between the S/C, to measure distances between the pairs of test masses, that occur in response to the passage of a gravitational wave. While the equations below apply to a generic GW signal from any astrophysical source, the coupling of the GW to the laser in this paper is meant for a frequency-stable GW signal as for the verification binary (VB).

The effect of a plane GW wave propagating far from the (Galactic binary) source in the vicinity of a GW detector where freely falling test masses are exchanging electromagnetic (EM) wave is to affect the properties of the EM wave \citep{Kaufmann1970}, i.e. LISA laser, where the light propagation is effected by the consequences of the constraint equation:
\begin{equation}
\Box F_{\mu\nu} + 2 R_{\mu\nu\alpha\beta} F^{\alpha\beta} = 0.
\end{equation}
The $d'$Alembertian, $\Box$, applied to an EM tensor $F^{\alpha\beta}$, relates to that of the 4-dimensional Riemanian $R_{\mu\nu\alpha\beta}$, curved space-time fluctuations by a factor of 2 modulo a symmetry in two of its four indices. The role of the ambient GW is to refract the EM wave akin to the scintillation of the EM waves from an optical source that passes through the time-varying Earth’s atmosphere.
\begin{figure*}
    \begin{minipage}[htbp]{\textwidth}
        \includegraphics[width=\textwidth]{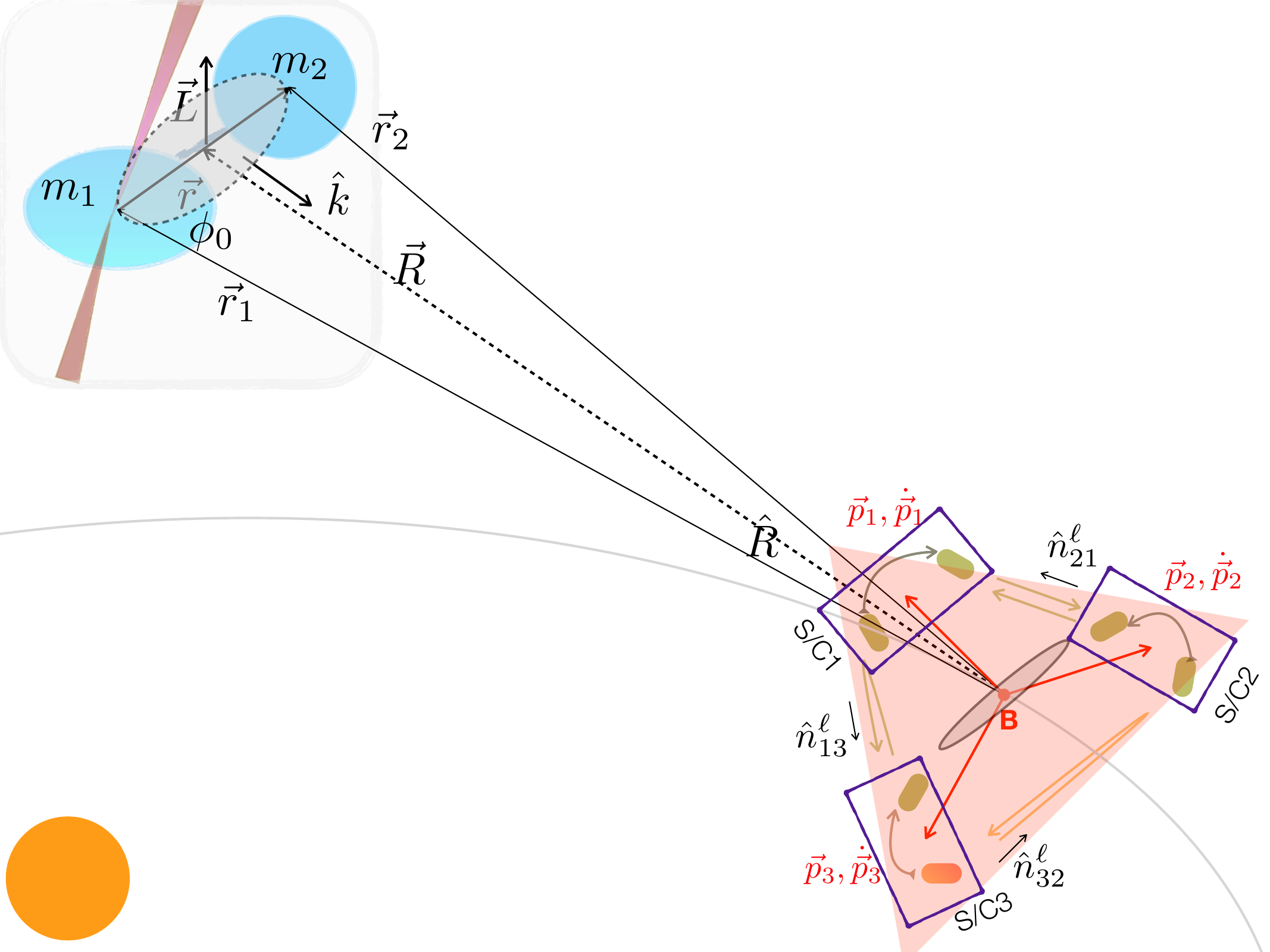}
        \caption{LISA inter spacecraft geometry and a VB binary with parameters and vectors defined in a coordinate system with the origin placed at the LISA Barycentre, labelled by B. Vectors $\vec{p}_i$ denote position(s) of S/C from B represented by red lines. Distances between the S/C for two directions are distinguished with $\hat{n}$ and the direction of a GW wave onto the LISA plane is represented by $\hat{k}$. Primary and secondary masses of binary are $m_1, m_2$ respectively with the position vectors $\vec r_1, \vec r_2$ from B. $\vec{R}$ is position vector to the centre of mass of the binary, whereas $\vec{r}$ defines the relative position of the masses on the orbital plane whose angular momentum is denoted by $\vec{L}$. The shaded red area is a plane spanned by LISA constellation plane which intersects with another 2-dimensional plane spanned by the projection of the binary orbital plane shown in grey circle. The complete set of interferometric measurements between locked lasers (for an exemplary laser frequency locking configuration) in all six optical benches are shown that will include GW signature (\ref{non_swap_min}) The orbit of LISA centroid around the Sun (in orange) is represented by grey curve.}
        \label{fig:coords}
    \end{minipage}
    \hfill
\end{figure*}
A weak field GW strain $h_{\mu\nu}$ and a linearized Riemann tensor, $R_{\mu\nu\alpha\beta}$, when inserted into the equation above give geodesics of an arbitrary massless particle between the test masses. Using Killing vectors, a mathematical tool that captures the symmetries of the spacetime, allowing us to understand how objects move within it 
\citep{Estabrook1975}, and the symmetry of the weak field gives the physical distance distortion caused by $h_{\mu\nu}$ affecting the photon flight time between the test masses. In other words, the doppler responses of a transverse traceless (TT) weak-field GW along a \textit{generic single} laser link and a \textit{singly transponded} 2-way laser link oriented in an arbitrary direction in 3-dimensional space (as shown in \cref{fig:coords}) is usually expressed in terms of its fractional frequency fluctuations, given by:
\begin{widetext}
\begin{eqnarray}
y^{\text{GW}}_{slr}(t_B) &=& \left[ 1 + \hat{k}^{\text{GW}}(t_B) \cdot \hat{n}^{\ell}_l(t_B) \right] \cross \frac{1}{2}\left\{ \Psi_l \left[ t_{\text{s}}(t_B) - \hat{k}^{\text{GW}}(t_B) \cdot \vec{p}_{\text{s}}(t_{\text{s}}(t_B)) \right] - \Psi_l \left[ t_B - \hat{k}^{\text{GW}}(t_B) \cdot \vec{p}_{\text{r}}(t_B) \right]  \right\} \label{eq:dopplerlink} \\ 
\Psi_{l}(t') &=& \frac{\hat{n}^{\ell}_l(t) \cdot \textbf{\text{h}}(t') \cdot \hat{n}^{\ell}_l(t)}{ 1 - \left[\hat{k}^{\text{GW}}(t) \cdot \hat{n}^{\ell}_l(t) \right]^2 }, \label{eq:geomlink}
\end{eqnarray}
\end{widetext}
where,
\begin{itemize}
  \item $t_B$ = time in an inertial reference frame filled by a plane GW, at LISA Barycentre
  \item $t$ = time in another inertial reference frame at Solar System Barycentre (SSB)
  \item $t'$ = proper time where the measurement is executed, for example, the clock's tick in a given S/C
  \item $t_{\mathrm{s}}$ = time coordinate of the S/C where laser beam is sent from
  \item $t_{\mathrm{r}}$ = time coordinate of the S/C where laser beam is received at
  \item $l$ = an index denoting armlength formed by laser propagation between two S/C
  \item $\hat{n}^{\ell}_l$ = unit vector of the laser beam $\ell$ along $l$
  \item $\hat{k}^{\text{GW}}$ = unit vector of the gravitational wave along the direction of its propagation in radiative zone
  \item $\vec{p}_{s}$ = position vector of S/C that is sending the laser beam with respect to a well defined origin
  \item $\vec{p}_{r}$ = position vector of S/C that is receiving with respect to the same origin
  \item $\textbf{\text{h}}$ (see Sec.~\ref{vbs}) is rank-2 tensor representing the culmination of two polarisations imprinted in the laser link $l$
\end{itemize}

Eq.~\ref{eq:dopplerlink} above adapted from from Eq.~1 in \citep{Vallis2005} is different by explicitly annotating the unsynchronised clocks to which the two distinct instances of the GW will impinge on the two ends of the link, making the above version more suitable to work with for the experimentalists. There are three distinct timescales at play here: (i) any event that occurs in radiative zone (i.e. far away from the source) (ii) some event that occurs at the retarded $t$ from the sending S/C and (iii) the version of the same event that occurs at the received $t$ at the receiving S/C. In the equation above, we have introduced  a coordinate frame at the LISA constellation centre, LISA barycentre, $t_B$, which is related to what is generally assumed in literature, the Solar System Barycentric (SSB) time, $t$. Observe that this $t$ is \textit{distinct} from those at $s$ or $r$.

In Eq.~\ref{eq:geomlink}, $\Psi_{l}$ is the projection of the inertial GW along its propagation direction onto a unit vector of laser propagating in between two ends of a theoretical detector consisting of two free-falling test masses \citep{Vallis2005}, where the recursive nature of the time is folded in. $y^{\mathrm{GW}}_{slr}$ is then the response of those test masses far enough apart by a distance such that the $\textbf{\text{h}}$ imprints itself twice onto the 2-mass detector: once at a retarded moment at $s$ and then again at a real time $t$, as measured by the receiving test-mass at $r$. 

The time, $t_{\mathrm{s}}$ is given by $t_{\mathrm{s}} = t_B - \abs {\vec p_r(t_B) - \vec p_s(t_s(t_B))}$, which is known as the light-propagation equation for non-relativistic length changes caused by GWs, which are essentially tidally deforming the 2-dimensional space along a length or, equivalently  changing the time taken to travel that distorted space due to GWs. Observe the \textit{recursive} nature of $t_B$ in its expression. The quantity $t_s(t_B))$ means a nested delay where $t_B$ depends on $t_s$ in the TDI terminology.
Alternatively, in terms of effective length, above mentioned equation can be re-casted as $L_l(t_B) = \abs{\vec p_r(t_B) - \vec p_s \left[ t_B - L_l(t_B)\right]}$, where this armlength is set from spatial point $r$ to another spatial point $s$ at emitted time, $t_B$.

\cref{fig:coords} shows the above-mentioned instrument parameters (in Eqs.~\ref{eq:dopplerlink},~\ref{eq:geomlink}) for the single-link response together with the GW parameters of a given binary (in Eqs.~\ref{eq:GWamplitude}) in a coordinate system with origin at the centre of LISA constellation (adapted from \citep{Wahlquist1987} and \citep{Heinzel2020}). The constellation as a whole orbits the Sun (represented by orange circle) trailing the Earth such that it circles the Sun in a ccw direction when viewed from the northern celestial pole. This coordinate system is different than the canonical Solar System Barycentre used in simulating LISA GW responses ~\cite[e.g.][]{Vallis2005, Petiteau08}. 

The VB parameters $m_1, m_2$ are the primary and secondary masses, where primary is taken to be the more massive one in this study. Their position vectors from LISA Barycentre are $\vec r_1, \vec r_2$ respectively, whereas $\vec R$ is the vector to the VB's centre-of-mass. Another intrinsic parameter, $\phi_0$, often ignored in previous analysis is the position of the $m_1$ relative to $m_2$ on the VB orbital plane represented by the grey shaded circle. The incoming nearly monochromatic GW signal's vector is represented by $\hat k$. The indexing conventions for the LISA constellation and its constituent test-masses, lasers amongst others are described in Sec.~\ref{non_swap_min} below. The unit vector $\hat n$ orienting the inter-S/C laser link is often used in exchange with the armlength $L_{ij}$ and Doppler link $D_{ij}$. The S/C motions, which will be significant in comparison to the GW signal are represented positions vectors and their rate of change by $\vec p_i, \dot \vec p_i$, respectively. 
\section{Frequencies in LISA lasers}\label{fplan}
LISA lasers will need to be controlled in their frequencies such that the interferometric beatnotes between those two lasers will need to fall in the range of $5-25$MHz \citep{Barke2015} as mentioned in Sec.~\ref{intro}.  This is necessitated by the fact that the three arms of LISA are time varying due to the individual S/C orbits\footnote{due to macroscopic push and pull of Sun, Earth, Moon, and Jupiter and Mars}, which causes Doppler shifts in the lasers propagating in-between a given pair of S/C. The $5-25$MHz range is driven by the requirement from the quadrant photodiodes \citep{Barranco2018} that register the beatnote between two interfered beams and the phasemeter \citep{Schwarze2019} that tracks the phase of the these beatnotes. Thus, the individual lasers should all be locked to a \textbf{primary} laser (PL) whose frequencies must have a determined offset between $5-25$MHz. Several hardware configurational solutions to achieve this requirement all described and derived in the ``\texttt{fplan} tool" document \citep{Heinzel2020}\footnote{internal to LISA Consortium}.
\begin{figure}[htb]
  \centering
    \includegraphics[width=0.45\textwidth]{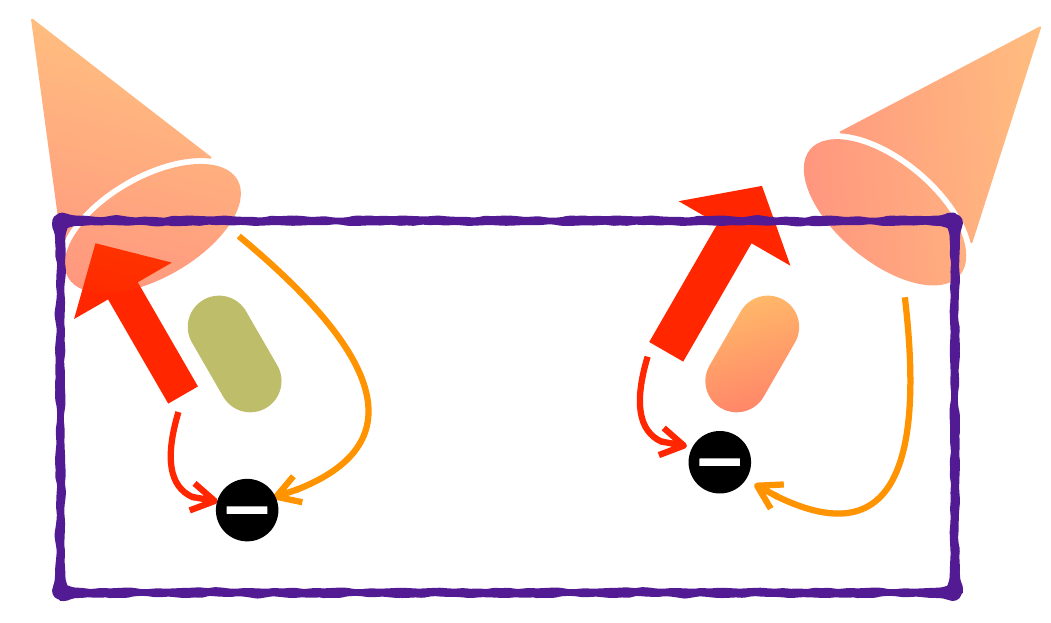}
  \caption{Configuration where the long arm signal is formed by interfering received orange beam from far S/C (represented by the rectangle) with local red beam originating in the \textit{same OB} where the interferometric signal (represented in the black and white cross) is measured. The red and green cylindrical objects represent the two local lasers on each optical bench (OB) in the S/C.\label{fig:nswap}}
\end{figure}
\begin{figure}[htb]
  \centering
    \includegraphics[width=0.45\textwidth]{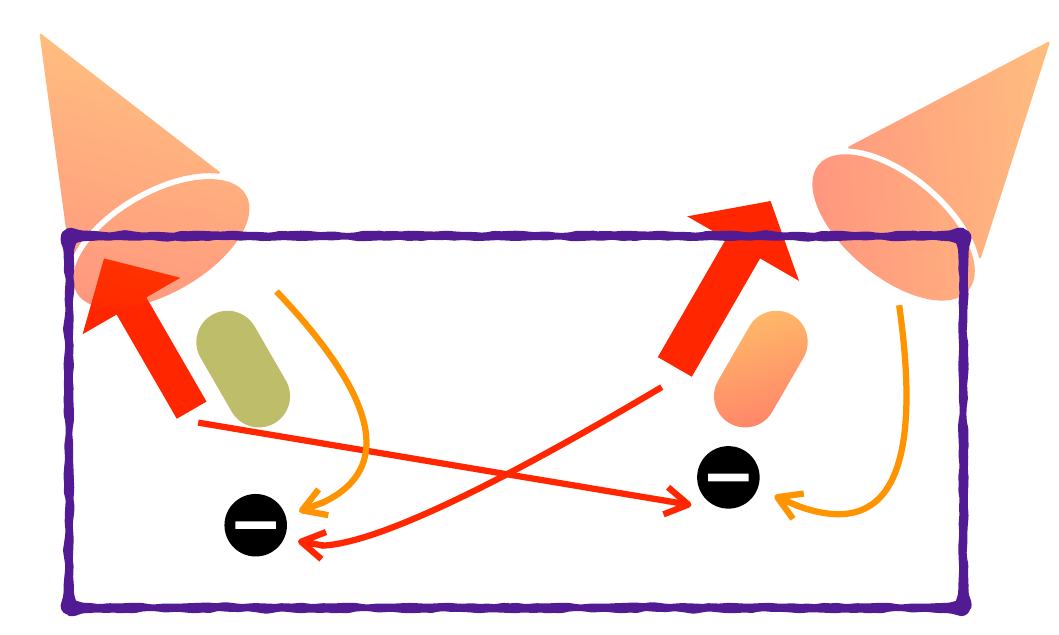}
  \caption{Similar to \cref{fig:nswap}, except the long arm signal is formed by interfering orange received beam from far S/C with local red beam originating in the \textit{adjacent OB}.\label{fig:swap}}
\end{figure}

In this paper, we consider only the configuration, shown in \cref{fig:nswap}, where the incoming beam from far S/C is interfered with the local laser on the OB where that long arm beatnote (also commonly known as the science signal) is measured. The incoming faint beam ($\sim 10^{-10}$\,Watts) is represented by the cone and the outgoing beam ($\sim$\,Watt) is represented by red arrow. There are two outgoing beams for each optical bench (OB) within the S/C, where the S/C is represented by the quadrilateral and the two rounded rectangles are the two lasers with offset phase lock between them. For this case, there exist six independent configurations of laser frequency offset locks where one of the 6 lasers is the primary laser and the remaining five are the secondary lasers (SL) locked to the former with predetermined offsets that keep the beatnotes within the boundaries of $5-25$ MHz for the duration of the observation of the mission, 4-10 years. Note, that the offsets do not need to be in the above-mentioned range for the beatnotes. Since the PL can be placed on any of the six OB, the total number of formations for \textit{non-swap}ping lasers yield thirty-six distinct choices of configurations.

In an alternative set-up, known as the \textit{frequency-swap} (\cref{fig:swap}), the long arm beatnote signal is performed by interfering received laser beam with the adjacent OB's local laser. This case will be studied in forthcoming paper and therefore will not be discussed in this study. We note that this setup, however, provides a staggering choices of seventy-two distinct ways to have any two lasers locked across the constellation which preserves the phasemeter and photodiode constraints mentioned above and should be studied for future LISA-type detectors.

In three of the six non-swap locking configurations, there are total of four single laser beam paths constituting the 2.5 million km long arm that are used to form a set of four long-arm measurements\footnote{\textit{classically} known as science interferometers, which there are six of for six free-running lasers across the constellation} with the GW-induced Doppler shift(s) collected along one or all three of the arms. In the remaining three locking configurations, there are total of three single laser beam paths constituting the 2.5\,Mkm long arm that form only three of such long-arm measurements with the fourth one being the local reference interferometer within the S/C. 

However, for all of the six non-swap configurations there are in total four main \footnote{carrier-carrier, there are also upper and lower side-band measurements in order to reduce clock jitter noise, see \citep{Tinto2018}} 
interferometric measurements which have GW responses, the later three mentioned above have that in the local interferometer as a fourth measurement described below. An example for each type are discussed in the subsections below. The starting point of this study is the result of \texttt{fplan} \citep{Heinzel2020}, where the final configuration for each non-swap configuration provides the relation of the dependence of the five secondary lasers on the primary laser with five distinct offsets, the laser frequency fluctuations and the Doppler shifts long the 3 arms while holding the boundary condition of $\pm5\mp25$MHz heterodyne frequencies. In this study we consider the configurations where the primary laser is on S/C 3 only.

The two examples of non-swap configurations can be understood as the algebraic manipulations required to determine the set of laser locking offsets. These offsets are applied to the five secondary lasers in relation to the primary laser (located in S/C3). Over a span of 10 years of S/C orbits, these adjustments ensure that the beatnotes formed by these lasers remain within the desired 5-25MHz range, and thus preventing a zero-crossing in the phasemeter. These applied offsets are distinct from the astrophysical Doppler shifts between the S/Cs that would for eg., be caused by the VBs made of white-dwarves. 
\subsection{Minimally locked non-swap configuration} \label{non_swap_min}
\begin{figure}[h!]
    \includegraphics[width=0.5\textwidth]{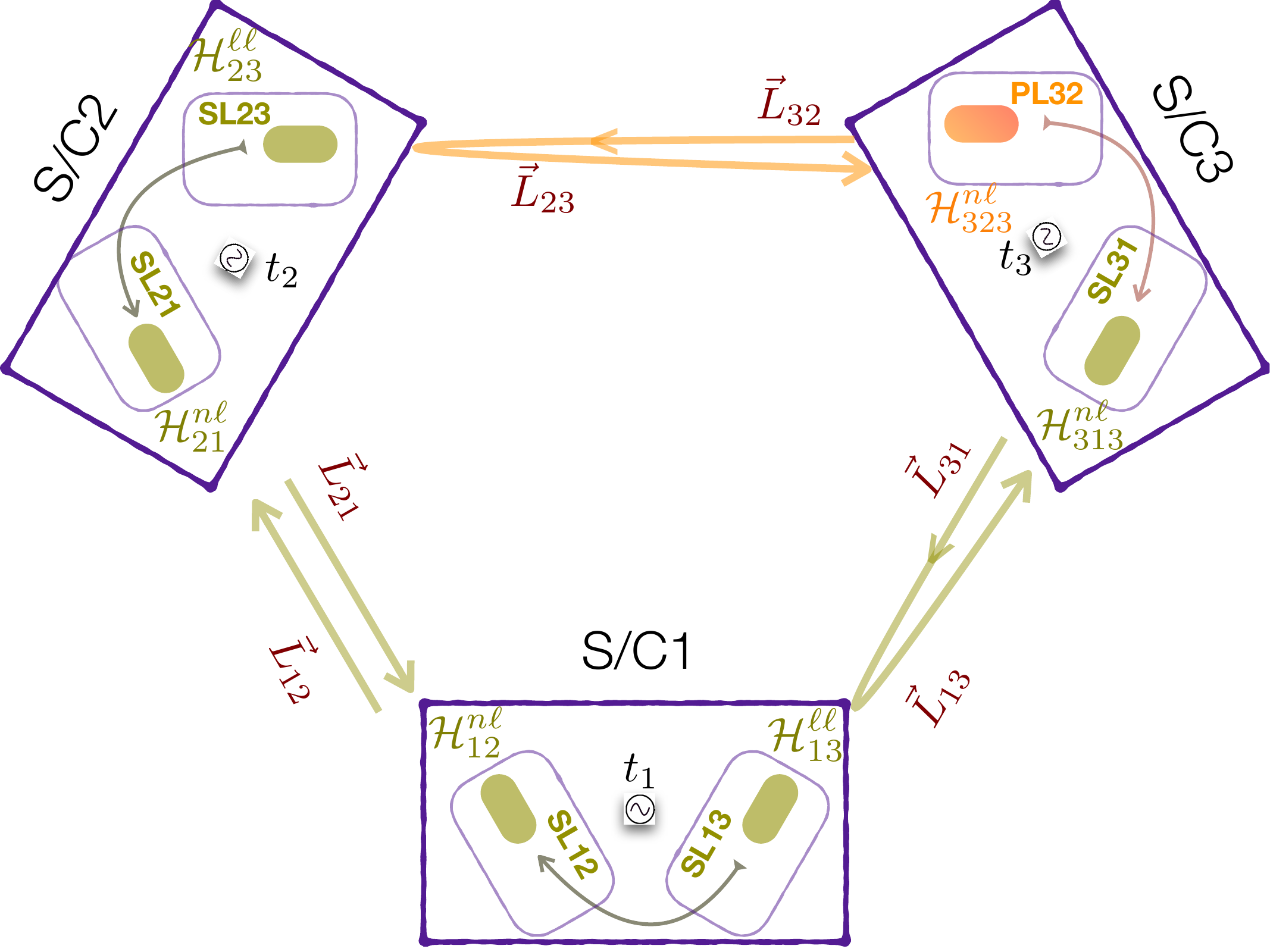}
  \caption{Laser locking configuration with minimum number of long arm Doppler shift lock to primary laser, PL32 (in orange) in S/C3 named, `N1c' (adapted from \citep{Heinzel2020}) in forming any of the four beatnotes, i.e. the maximum number of $L$ used to lock the secondary laser is one. There are four distinct long-arm measurements across the constellation shown by the arrows, labelled by $\mathcal{H}$. The local interferometers within the S/C (Eq.~\ref{eq:nswap1c:locrefs}) are not shown. The armlengths (including Doppler shifts) are denoted by $\vec L_{ij}$ indicated by the lines connecting the S/C with arrows. The arrow connection any two lasers within the S/C shows the locking between them. \label{fig:n1c}}
\end{figure}

The heterodyne beatnotes' frequencies, $\mathcal{H}$\footnote{inspired by GW nomenclature, strain, $h\sim \delta{L}/L$, where L is the armlength. The beatnotes formed by laser beams propagated in the long arms will capture GW signature of a VB.}, are expressed in terms of the primary laser frequency $f_0^{\ell}$, a given set of laser frequency offsets $O_{k}$ for the configuration in \cref{fig:n1c}, and Doppler $D_{ij}$ shifts picked up along the long arm by the interchanging laser beams. In the figure, the primary laser (in orange), PL32 is chosen to be on S/C3.  The rest of the 5 lasers (in green) are secondary: SL23, SL21, SL12, SL13 and SL31. The adjacent SL31 is locally locked to PL32. SL23 is locked to PL32 with a single Doppler shift along the arm $\vec L_{32}$. SL21 is locally locked to SL23. As SL21, SL12 is locally locked to SL13, while SL13 is locked to SL31 with a single Doppler shift along arm $\vec L_{31}$. This particular setup of locking scheme is called `N1c'. 

The label and names for LISA are mostly taken from previous notation conventions, see eg.\cite{Hewitson2021, Bayle2023}. In this paper, $\mathcal{H}$ has an unconventional label denoted by 2 or 3 S/C indices. The upper index in $\mathcal{H}$ is referred to $\ell\ell$ if that beatnote is formed by locking to the local laser, i.e. including the long-arm Doppler shift such that there is no trace of the physical shift caused by long arm propagation in that beatnote. However, it is referred to $n\ell$ where the beatnote preserves that physical shift, implying that only $\mathcal{H}^{n\ell}$ term has the Doppler shift from a long-duration astrophysical source.
The lower index in $\mathcal{H}$ denotes the laser beam path(s) of sometimes two lasers that are used to form the beatnote and other times of a common laser that is used to form the beatnote. For instance, $\mathcal{H}_{12}$ is formed by interfering lasers SL12 and SL21, where the beams from two physically distinct lasers are exchanged after having travelled all three armlengths (Eq.~\ref{eq:nswap1c:h12}). Whereas, $\mathcal{H}_{313}$ is formed by SL31 interfered with itself after a round-trip armlength travel, where SL13 is used to send back fresh beam after being locked to the frequency of SL31 defining the meaning of the term \textit{transponded} beam (Eq.~\ref{eq:nswap1c:h313}). The topological view of locking schemes in `N1c' is shown in \cref{fig:n1c}, where the reference and/or test-mass interferometers (Eq.~\ref{eq:nswap1c:locrefs}) are not shown. We consider only the six long arm carrier-to-carrier beatnotes in this study. The three spacecrafts are labelled S/C1, S/C2 and S/C3 respectively. The arrows connecting the two lasers within a S/C represents the locking direction, such that in S/C3, the secondary laser, SL31 is locked in its phase with PL32; in S/C2, the secondary laser SL21 is locked to the phase of SL23 and finally in S/C1, the secondary laser of SL12 is locked to the phase of SL13.

Thus it turns out that in order to keep to the phasemeter condition, and for the `N1c' configuration we have, a total of 2 distinct round-trip long-arm measurements made by interfering with an identical laser and 2 distinct one-way long-arm measurements made by interfering two distinct lasers. We highlight that this setup is completely different than the case with free-running lasers considered widely in literature for TDI analysis and is further illustrated in the subsections below for `N1c'\footnote{The names `N1c' and `N2a' are taken from  \citep{Heinzel2020}} and for a second configuration named `N2a' in Sec.~\ref{non_swap_max} that has another set of laser offset frequencies. 

For the configuration in \cref{fig:n1c}, the resulting reference and long-arm beatnotes are (adapted from \cite{Heinzel2020}):
    \begin{subequations}
    \label{eq:nswap1c}
        \begin{align}
        \mathcal{H}_{11} &=O5 \;\;\;\; \mathcal{H}_{22} = O4 \;\;\;\; \mathcal{H}_{33} = -O3,\label{eq:nswap1c:locrefs}\\
        \mathcal{H}^{\ell\ell}_{23} &= -O1 \;\;\;\; \mathcal{H}^{\ell\ell}_{13} = -O2,\label{eq:nswap1c:h23h13} \\
        \mathcal{H}^{n\ell}_{12} &= -f_{0,[31]} + f_{0,[21][32]} \nonumber \\
        & + D_{21} + D_{32,[21]} - D_{31} \nonumber \\
        & + O4_{,[21]} + O1_{,[21]} - O5 - O2 - O3_{,[31]}, \label{eq:nswap1c:h12}\\
        \mathcal{H}^{n\ell}_{21} &= -f_{0,[32]} + f_{0,[12][31]} \nonumber \\
        & - D_{32} + D_{12} + D_{31,[12]} \nonumber \\
        & + O5_{,[12]} + O2_{,[12]} + O3_{,[12][31]} - O1 - O4, \label{eq:nswap1c:h21}\\
        \mathcal{H}^{n\ell}_{313} &= -f_0 + f_{0,[13][31]}  \nonumber \\
        &+ D_{13} + D_{31,[13]} \nonumber \\
        &+ O2_{,[13]} - O3+ O3_{,[13][31]}, \label{eq:nswap1c:h313}\\
        \mathcal{H}^{n\ell}_{323} &= -f_0 + f_{0,[23][32]}  \nonumber \\
        &+ D_{32,[23]} + D_{23} \nonumber \\
        &+ O1_{,[23]}, \label{eq:nswap1c:h323}
        \end{align}
    \end{subequations}
where,
\begin{itemize}
  \item Eqs.~\ref{eq:nswap1c:locrefs} are the measurements of the local interferometers measuring the phase difference between two adjacent lasers on the two OB.
  \item Eqs.~\ref{eq:nswap1c:h23h13} are the measurements of the long arm interferometers measuring the phase difference between two lasers on board two S/C without Doppler shifts in them, see \cref{fig:n1c}.
  \item Eqs.~\ref{eq:nswap1c:h12}, ~\ref{eq:nswap1c:h21}, ~\ref{eq:nswap1c:h313}, ~\ref{eq:nswap1c:h323} are the measurements of the long arm interferometers measuring the phase difference between two lasers on board two S/C with varying number of  Doppler shifts in them, see \cref{fig:n1c}.
  \item $f_{0}^{\ell}$ is the instantaneous laser frequency of the primary laser delayed by one link, measured according to clock in S/C1
  \item All two-index subscripts inside square brackets $f_{0,[ij]}^{\ell}, O_{,[ij]}, D_{[ij]}$ are physically occurring delays along the armlength $\vec L_{ij}$.
  \item $D$ is Doppler shift at the time of arrival at the receiver.
  \item $O1 ... O5$ are sets of five distinct laser frequency offsets whose different linear combinations are applied to individual secondary lasers, see \cref{fig:n1c_B12}.
\end{itemize}
The equations above do not specify any reference frame where the time is referenced with respect to. For $\mathcal{H}^{n\ell}_{12}$, adding a reference clock to which its digitised time-series will be referred to, we get:
\begin{eqnarray}
\mathcal{H}^{n\ell}_{12}(t_1) 
    &=& -f_{0,[31]}^{\ell}(t_1) + f_{0,[21][32]}^{\ell}(t_1) \nonumber \\
    &+& D_{32,[21]}^{\text{VB}}(t_1) - D_{31}^{\text{VB}}(t_1) + D_{21}^{\text{VB}}(t_1) \nonumber \\
    &-& O1_{,[21]}(t_1) - O2(t_1) + O3_{,[31]}(t_1) \nonumber \\ 
    &+& O4_{,[21]}(t_1) - O5(t_1),\label{eq:nswap_h12_VBdoppler}
\end{eqnarray}
where,
\begin{itemize}
  \item $\mathcal{H}^{n\ell}_{12}(t_1)$ is the beatnote signal measured on OB12 in S/C1 with respect to clock 1\footnote{Each S/C has a one primary (oscillatory) \textbf{clock} and one primary (digital) \textbf{timer} to which all measurements are referred to.}, $t_1$ (labelled by USO1 in \cref{fig:n1c_B12}.
  \item $f_{0,[31]}^{\ell}$ is the laser frequency noise of the primary laser delayed by one link, measured according to clock in S/C1.
  \item $D^{\text{VB}}_{ij}$ is Doppler shift from a VB at the time of arrival at the receiver, $j$.
  \item $O1$\footnote{$O$ is used to represent the residual noise from laser lock, which is typically referred to as `laser offset'}, is laser offset applied to SL23; $O1+O4$ are two distinct laser offsets applied to SL21; $O3+O2+O5$ are three distinct laser offsets applied to SL12 frequencies; and $O3+O2$ are two distinct laser offsets applied to SL13.
\end{itemize}

In the Eq.~\ref{eq:nswap_h12_VBdoppler} above, we have ignored all Doppler shifts arising from different parts of the instrument and side-band modulations, etc., isolating the only contribution from VB, hence the label, $D^{\text{VB}}$, while contributions from the laser frequency fluctuations, laser frequency offsets are included in the overall $\mathcal{H}$ expression. Furthermore, we exclude test-mass interferometers, such that there are only total of six carrier-to-carrier beatnote signals for the long arm and reference interferometers. Additionally, many noise terms related to these interferometers are ignored in this paper, for eg. clock noise \citep{Hartwig2021}, tilt-to-length noise (Wanner et al 2023, in prep), the many contributions from the gravitational reference system \citep{GRScqg2016}, and due to other control loops. 

\textit{Note} that the label $D_{ij}$ in the equations and $L_{ij}$ in the figures are used interchangeably. In literature $L_{ij}$ is often used to mean the nominal armlength in the absence of any external forces and $D_{ij}$ is then the added variation due to the forces. However, for the purpose of this paper we use both the terms to mean similar quantity where variations due to any external force is included. Additionally, while $D$ is used to mean Doppler shifts collected along the arm (owing to various processes), $L$ is used to mean the armlength. Despite these different meanings, the $D$ in the equations can be exchanged with $L$ in the figure and vice-versa.
\subsection{Breakdown of individual beam paths for the minimally locked beatnotes}
\label{n1c_resp}
In this subsection we show the laser paths used to build the interferometric signals explicitly and the coupling of a verification binary GW signal to it for their parameters from \ref{vbs} for the laser offset locking configuration `N1c'. Consider the measurement made at OB12 where $\mathcal{H}^{n\ell}_{12}$ (as shown in \cref{fig:n1c_B12}) is the beatnote signal formed by interfering the local beam in S/C1, SL12, and the distant beam from S/C2, SL21, with a delay from that S/C. This and all other interferometrer equations in Eqs.~\ref{eq:nswap1c} use the convention where the the local beam is subtracted from the distant beam. As explained above, the beatnotes that contain the long arm Doppler shifts are constructed using laser beams that travel across the constellation atleast two arms before being interfered with another laser beam which has also possibly travelled across one or two arms. For $\mathcal{H}^{n\ell}_{12}$ (in Eq.~\ref{eq:nswap1c:h12}) can be derived as (dropping the temporal $t_i$ reference):
\begin{figure}[htb]
  \centering
    \includegraphics[width=0.45\textwidth]{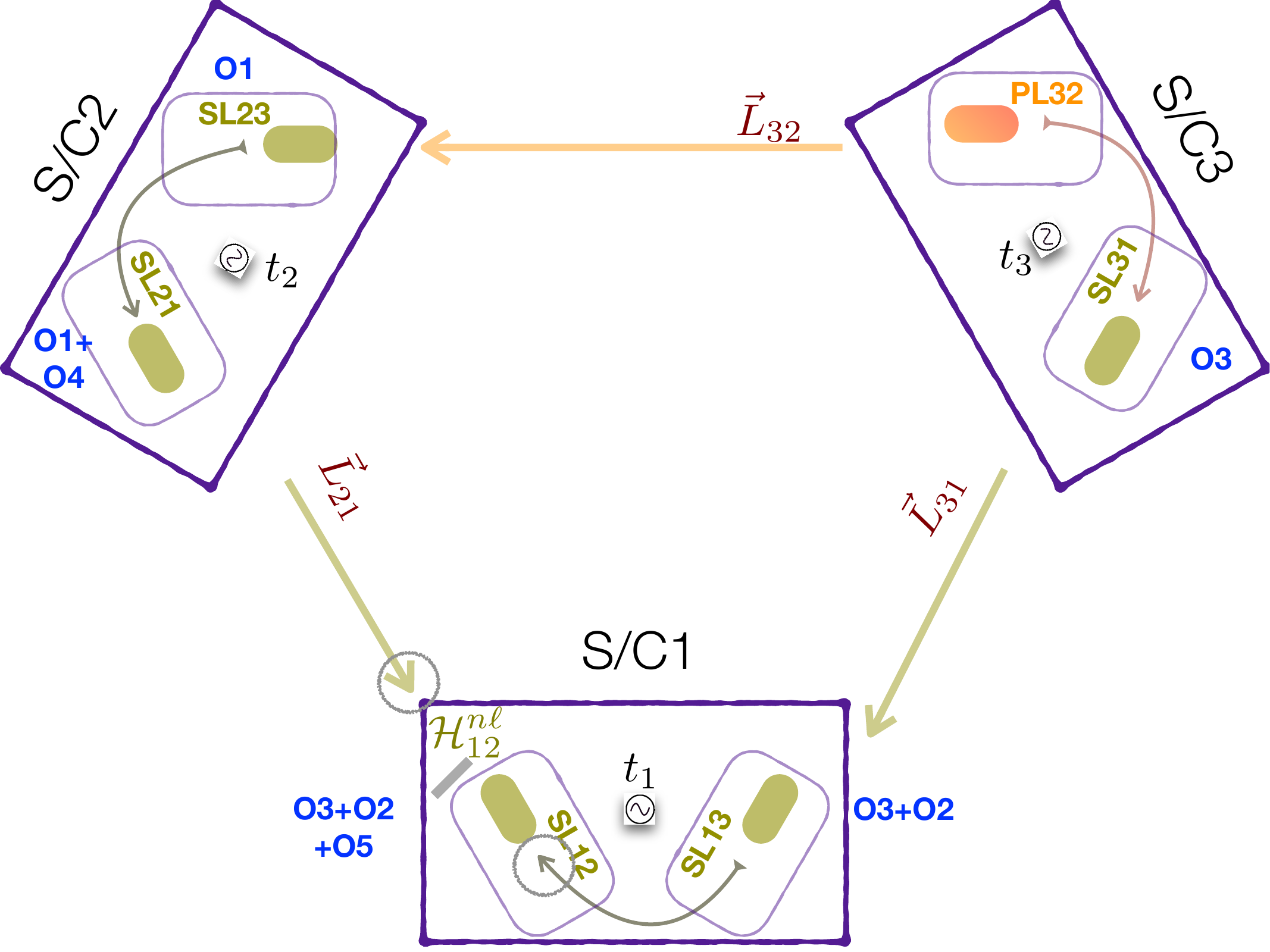}
  \caption{A pair of laser beams showing their respective paths along the long-arm (indicated by grey circles) in forming the interferometric measurement $\mathcal{H}^{n\ell}_{12}$ (see, Eq.~\ref{eq:nswap_h12_beam_paths}) where the interference is indicated by slanted grey line at OB12. This is an example of a long-arm interferometric measurement, where two distinct one-way beams are subtracted from one another. USO are time-references, Ultra Stable Oscillators for S/C USO1, USO2, and USO3 as labelled. \label{fig:n1c_B12}}
\end{figure}
\begin{figure}[htb]
  \centering
    \includegraphics[width=0.45\textwidth]{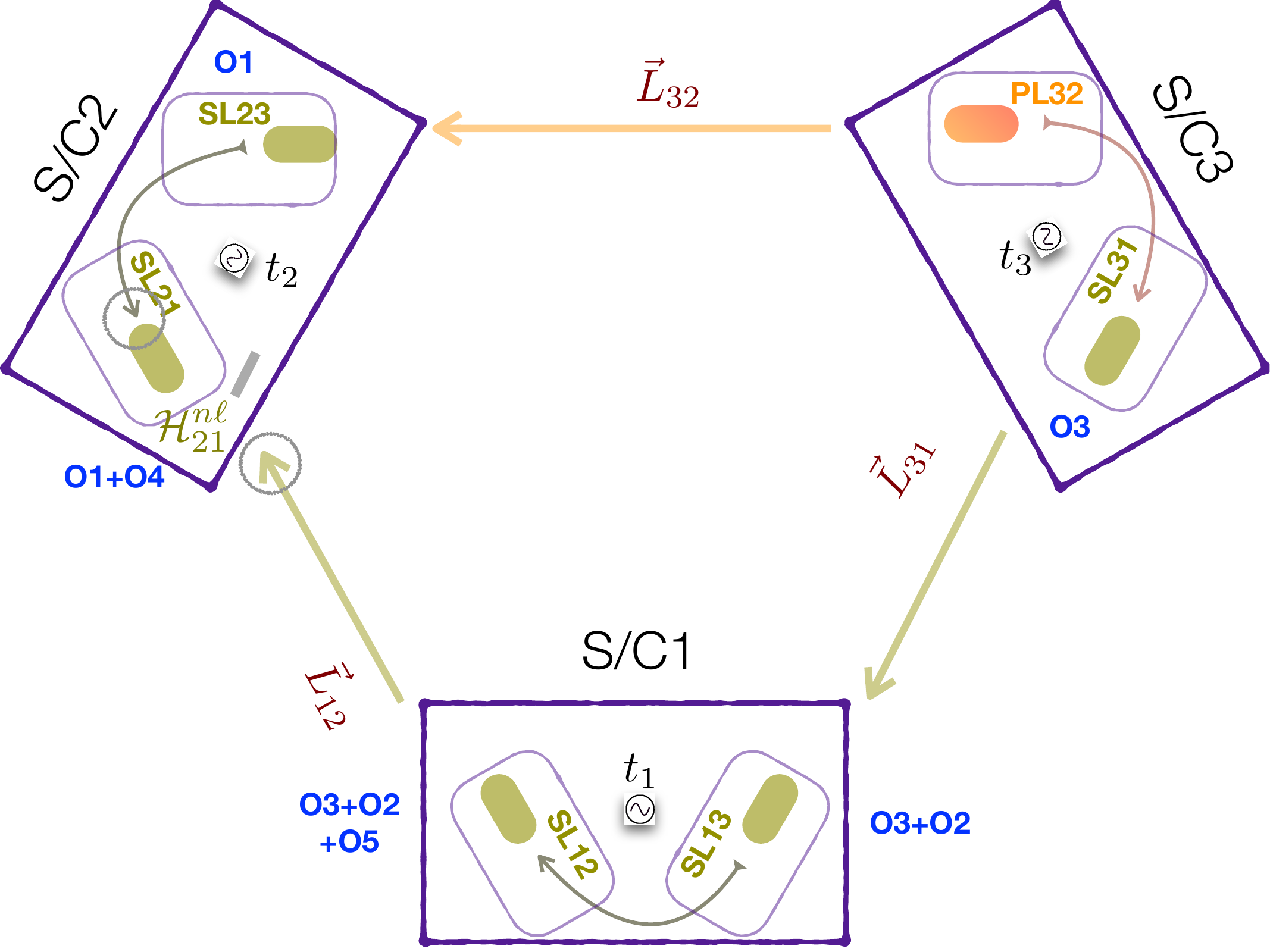}
  \caption{Similar to \cref{fig:n1c_B12}, the laser beam paths to form the interferometric measurement $\mathcal{H}^{n\ell}_{21}$ (see, Eq.~\ref{eq:nswap1c:h21}) Here, the laser from distant S/C1 (OB12), i.e. SL12 is interfered with local beam (SL21) on OB21.\label{fig:n1c_B21}}
\end{figure}
\begin{figure}[htb]
  \centering
    \includegraphics[width=0.45\textwidth]{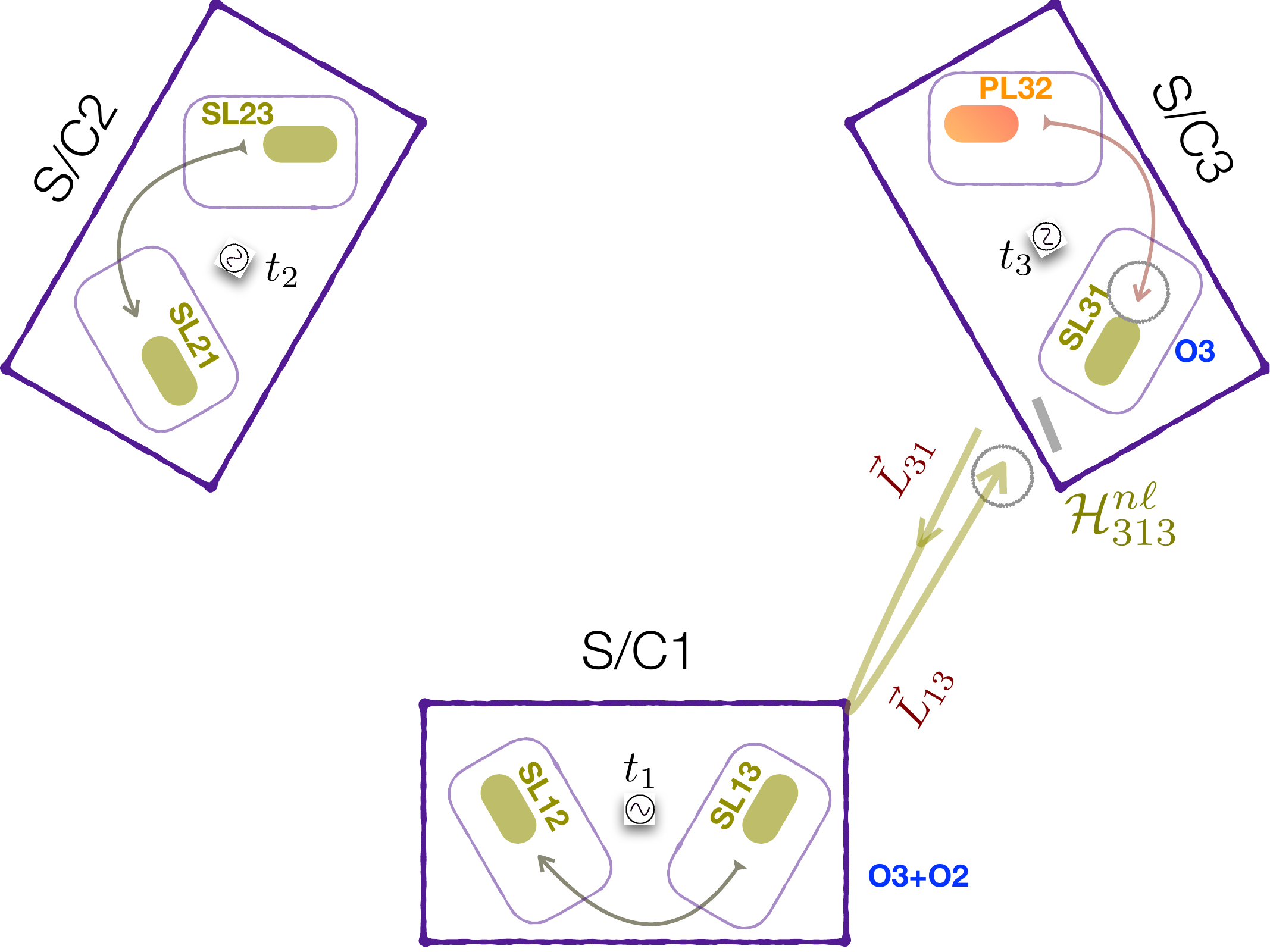}
  \caption{The laser beam paths to form the interferometric measurement $\mathcal{H}^{n\ell}_{313}$ (see, Eq.~\ref{eq:nswap1c:h313}). Unlike in \cref{fig:n1c_B21} and, \cref{fig:n1c_B12}, this is a measurement in the OB31, where a local laser on OB31 in S/C3, i.e. PL32, is interfered with the same beam transponded back (a two-way measurement) from OB13 from S/C1.\label{fig:n1c_B313}}
\end{figure}
\begin{figure}[h!]
  \centering
    \includegraphics[width=0.45\textwidth]{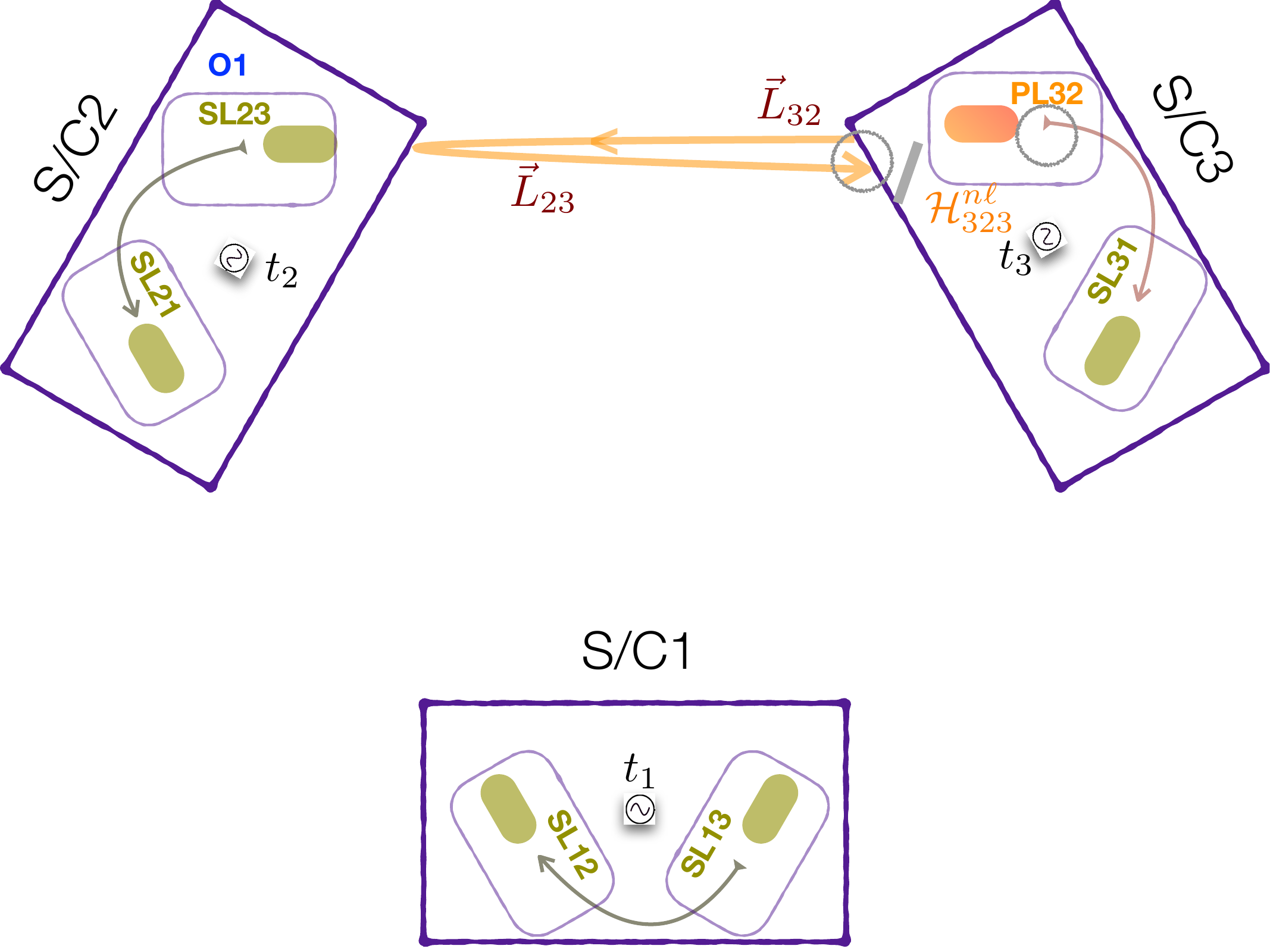}
  \caption{Similar to \cref{fig:n1c_B313}, the laser beam paths to form the interferometric measurement $\mathcal{H}^{n\ell}_{323}$ (see, Eq.~\ref{eq:nswap1c:h323}). This is a measurement in the OB23, where a local laser on OB31 in S/C3, i.e. PL32, is interfered with the same beam transponded back (a two-way measurement) from OB23 from S/C1 .\label{fig:n1c_B323}}
\end{figure}
\begin{eqnarray}
\mathcal{H}^{n\ell}_{12} &=& [T_{21}(\text{SL}21)] \;\;\;\;\;\;\;\;\;\;\;\;\; - [\text{SL}12] \nonumber \\
&=& [T_{21}(T_{32}(\text{PL}32))] \;\;\;\;\; - T_{31}([\text{PL}32]) \nonumber \\
&=& [T_{21}(T_{32}(f_{0}^{\ell}))] \;\;\;\;\;\;\;\;\;\: - [T_{31}(f_{0}^{\ell})] \nonumber \\
&& [+ T_{21}(D_{32}^{\text{VB}}) + D_{21}^{\text{VB}}] \;\; - [D_{31}^{\text{VB}}] \nonumber \\
&& [+ T_{21}(O1+O4)] - [O5 + T_{31}(O3) + O2] \label{eq:nswap_h12_beam_paths}
\end{eqnarray}
where,
\begin{itemize}
  \item $\mathcal{H}^{n\ell}_{12}, f_{0}^{\ell}, O1, ... O5$ are described below Eqs.~\ref{eq:nswap1c},~\ref{eq:nswap_h12_VBdoppler}.
  \item $D_{ij}^{\text{VB}}$ is Doppler shift of the VB signal at the time of arrival at the receiver, $j$.
  \item $T_{ij}$ is delay operator for light travel time along the arm travelling from $i \rightarrow j$, note that this is written alternatively in form of $[ij]$ as subscript, for eg. in Eqs.\ref{eq:nswap1c:h12}, \ref{eq:nswap1c:h21}, \ref{eq:nswap1c:h313}, \ref{eq:nswap1c:h323}, \ref{eq:nswap_h12_VBdoppler} and elsewhere.
\end{itemize}

Observe in Eq.~\ref{eq:nswap_h12_beam_paths} above, that the number of Doppler shifts (3 in the example above) in the phasemeter output is different than the number of laser offsets (5 in the example above). The number of Doppler shifts is also distinct from the number of instances of the instantaneous laser frequency noise $f^{\ell}_0$. However, terms with $f^{\ell}_0$ always appear in two for all of the phasemeter time-series in Eqs.~\ref{eq:nswap1c}, ~\ref{eq:nswap_h12_VBdoppler} and ~\ref{eq:nswap2a}, ~\ref{eq:nswap_h212_n2a_beam_paths} below and in fact for all non-swap configurations, albeit with different delays.

As mentioned above, in the minimally-locked configuration in \cref{fig:n1c}, there are a total of four distinct-single beam paths originating from the primary laser that are interfered with a local laser in constructing those four distinct beatnotes respectively, which are shown in Figs.\ref{fig:n1c_B12},\ref{fig:n1c_B21},\ref{fig:n1c_B313},\ref{fig:n1c_B323}. Thus, the \textit{single arm} Doppler response stated in Eq.~\ref{eq:dopplerlink} coupling to each of the beatnote signals in Eqs.~\ref{eq:nswap1c:h12}, \ref{eq:nswap1c:h21}, \ref{eq:nswap1c:h313} and \ref{eq:nswap1c:h323} is \textit{fundamentally different} than for the configuration with free-running lasers widely used in literature. For the above mentioned example, $\mathcal{H}^{n\ell}_{12}$ the two beams used to construct it are shown in \cref{fig:n1c_B12} and expressed in Eq.~\ref{eq:nswap_h12_beam_paths} corresponding to the two single-path Doppler links. Concretely they are one beam with the following path: PL32 $\rightarrow$ SL23 $\rightarrow$ SL21 $\rightarrow$ SL12 and another beam with the following path: PL32 $\rightarrow$ SL31 $\rightarrow$ SL13 $\rightarrow$ SL12, which are interfered (subtracted) to form the beatnote signal at OB12, where at each secondary laser, a set of offsets are applied as labelled in the \cref{fig:n1c_B12}.

The three individual Doppler terms $D$ terms are equivalent to the nomenclatures of fractional frequency deviations $y$ and instantaneous GW signature from Eqs.~\ref{eq:dopplerlink},~\ref{eq:geomlink} respectively, such that $y_{slr} = D_{slr}$, both referred to coordinate system at the LISA Barycentre with time coordinate, $t_B$. We often suppress the $l$ in D by writing it as $ = D_{sr}$. In Eq.~\ref{eq:nswap_h12_beam_paths}, suppressing the offsets and laser frequency noise, OB12 registers the following \textit{six} distinct impinges of the same GW signal along different orientations of the arms, with the clock reference of USO1, $t_1$, as:
\begin{widetext}
\begin{eqnarray}
\mathcal{H}^{n\ell,\text{GW}}_{12}(t_1) &=& \left[ T_{21}\left( y^{\text{GW}}_{3[32]2}(t_B) \right) + y^{\text{GW}}_{2[21]1}(t_B)\right] - \left[y^{\text{GW}}_{3[31]1}(t_B)\right] \nonumber \\
&=& T_{21}\left( \left[ 1 + \hat{k}^{\text{GW}}(t_B) \cdot \hat{n}_{[32]}^{\ell}(t_B) \right] \frac{1}{2} \cross \left\{ \Psi_{[32]} \left[ t_3(t_B) - \hat{k}^{\text{GW}}(t_B) \cdot \vec{p}_3(t_3(t_B))\right] - \Psi_{[32]} \left[ t_B - \hat{k}^{\text{GW}}(t_B) \cdot \vec{p}_2(t_B) \right]  \right\} \right) \nonumber \\
&+& \left[ 1 + \hat{k}^{\text{GW}}(t_B) \cdot \hat{n}_{[21]}^{\ell}(t_B) \right] \frac{1}{2} \cross \left\{ \Psi_{[21]} \left[ t_2(t_B) - \hat{k}^{\text{GW}}(t_B) \cdot \vec{p}_2(t_2(t_B))\right] - \Psi_{[21]} \left[ t_B - \hat{k}^{\text{GW}}(t_B) \cdot \vec{p}_1(t_B) \right]  \right\} \nonumber \\
&-& \left[ 1 + \hat{k}^{\text{GW}}(t_B) \cdot \hat{n}_{[31]}^{\ell}(t_B) \right] \frac{1}{2} \cross \left\{ \Psi_{[31]} \left[ t_3(t_B) - \hat{k}^{\text{GW}}(t_B) \cdot \vec{p}_3(t_3(t_B))\right] - \Psi_{[31]} \left[ t_B - \hat{k}^{\text{GW}}(t_B) \cdot \vec{p}_1(t_B) \right]  \right\} \label{eq:n1cH12} 
\end{eqnarray}
\end{widetext}
where, 
\begin{itemize}
    \item $y^{\text{GW}}_{3[32]2}(t_B)$ is the 2-pulse response time-series in a Doppler link from OB32 to OB12.
    \item $y^{\text{GW}}_{2[21]1}(t_B)$ is the 2-pulse response time-series in a Doppler link from OB21 to OB12.
    \item $y^{\text{GW}}_{2[31]1}(t_B)$ is the 2-pulse response time-series in a Doppler link from OB31 to OB13
    \item $\Psi, \hat n, \hat k, \vec p_i, t_B$ are described under Eqs.~\ref{eq:dopplerlink},~\ref{eq:geomlink}.
    \item $t_1, t_2, t_3$ are time stamps from the three clocks in S/C1, S/C2, and S/C3 respectively (labelled by USO1, USO2 and USO3 in \cref{fig:n1c_B12}.
    \item $T_{21}$ is the physically occurring delay along the light beam travelling from optical bench OB21 to OB12.
\end{itemize}
Observe in Eq.~\ref{eq:n1cH12} above, for the S/C position vector terms $\vec p$, the mixing of the different time-frames referenced with three different clocks placed in the three S/C, where each clock in the given S/C is designed to drive all the frequencies that need referencing to a common (one main) clock \citep{Heinzel2020b}. This is indicated in \cref{fig:n1c_B12} by the three Ultra Stable Oscillators (USO): USO1, USO2, and USO3, the main clocks for the 3 S/C. These are further referred by the three corresponding digital timers, $t_1, t_2, t_3$ respectively, common, but unique to each S/C. The doppler shifts above in Eq.~\ref{eq:nswap_h12_VBdoppler} (for e.g. $D_{32,21}^{\text{VB}}(t_1)$) are given by the expressions in Eqs.~\ref{eq:dopplerlink}, and ~\ref{eq:geomlink} for a given signal of determined shape. These are written out for the beatnotes for two of the exemplary laser locking configurations in subsections below.

Similarly, the beatnote $\mathcal{H}^{n\ell}_{21}$ is measured on optical bench, OB21 which is obtained by subtracting the beam with the path: PL32 $\rightarrow$ SL23 $\rightarrow$ SL21 from another beam with the path: PL32 $\rightarrow$ SL31 $\rightarrow$ SL13 $\rightarrow$ SL12 to form the beatnote signal, with the clock reference of USO2, $t_2$ (as shown in \cref{fig:n1c_B21}) as:  
\begin{widetext}
\begin{eqnarray}
\mathcal{H}^{n\ell,\text{GW}}_{21}(t_2) &=& \left[ y^{\text{GW}}_{3[32]2}(t_B)\right] - \left[ T_{12}\left( y^{\text{GW}}_{3[31]1}(t_B) \right) + y^{\text{GW}}_{1[12]2}(t_B) \right] \nonumber \\
&=& \left[ 1 + \hat{k}^{\text{GW}}(t_B) \cdot \hat{n}_{[32]}^{\ell}(t_B) \right] \frac{1}{2} \cross \left\{ \Psi_{[32]} \left[ t_3(t_B) - \hat{k}^{\text{GW}}(t_B) \cdot \vec{p}_3(t_3(t_B))\right] - \Psi_{[32]} \left[ t_B - \hat{k}^{\text{GW}}(t_B) \cdot \vec{p}_2(t_B) \right]  \right\} \nonumber \\
&-& T_{12}\left( \left[ 1 + \hat{k}^{\text{GW}}(t_B) \cdot \hat{n}_{[31]}^{\ell}(t_B) \right] \frac{1}{2} \cross \left\{ \Psi_{[31]} \left[ t_3(t_B) - \hat{k}^{\text{GW}}(t_B) \cdot \vec{p}_3(t_3(t_B))\right] - \Psi_{[31]} \left[ t_B - \hat{k}^{\text{GW}}(t_B) \cdot \vec{p}_1(t_B) \right] \right\} \right) \nonumber \\
&-& \left[ 1 + \hat{k}^{\text{GW}}(t_B) \cdot \hat{n}_{[12]}^{\ell}(t_B) \right] \frac{1}{2} \cross \left\{ \Psi_{[12]} \left[ t_1(t_B) - \hat{k}^{\text{GW}}(t_B) \cdot \vec{p}_1(t_1(t_B))\right] - \Psi_{[12]} \left[ t_B - \hat{k}^{\text{GW}}(t_B) \cdot \vec{p}_2(t_B) \right]  \right\} \label{eq:n1cH21}
\end{eqnarray}
\end{widetext}
where,
\begin{itemize}
    \item $y^{\text{GW}}_{3[32]2}(t_B)$ is described below Eq.~\ref{eq:n1cH12}.
    \item $y^{\text{GW}}_{3[31]1}(t_B)$ is the 2-pulse response time-series in a Doppler link from OB31 to OB13.
    \item $y^{\text{GW}}_{1[12]2}(t_B)$ is the 2-pulse response time-series in a Doppler link from OB12 to OB21.
    \item $\Psi, \hat n, \hat k, \vec p_i, t_B$ are described under Eqs.~\ref{eq:dopplerlink},~\ref{eq:geomlink}.
    \item $t_1, t_3$ are time stamps from the two clocks in S/C1 and S/C3 respectively.
    \item $T_{12}$ is the physically occurring delay along the light beam travelling from optical bench OB12 to OB21.
\end{itemize}

As described previously, the beatnote $\mathcal{H}^{n\ell}_{313}$ is different in nature compared to those of  $\mathcal{H}^{n\ell}_{12}$ and $\mathcal{H}^{n\ell}_{21}$ derived above, where $\mathcal{H}^{n\ell}_{313}$ is formed by interfering a pair of two distinct laser beam paths from an identical physical laser, which is why it has a 3-index label to distinguish it from the the beatnotes, $\mathcal{H}^{n\ell}_{12},\mathcal{H}^{n\ell}_{21}$. For $\mathcal{H}^{n\ell}_{313}$, this physical laser is the primary laser, PL32. Concretely, this beatnote is measured by subtracting a two-way beam with the path: PL32 $\rightarrow$ SL31 $\rightarrow$ SL13 $\rightarrow$ SL31 from another beam which is located on OB32, that is the primary laser beam PL32, shown in \cref{fig:n1c_B313}). The resulting coupling to VB signals in the doppler links are given by:
\begin{widetext}
\begin{eqnarray}
\mathcal{H}^{n\ell,\text{GW}}_{313}(t_3) &=& \left[ T_{13}\left( y^{\mathrm{GW}}_{3[31]1}(t_B)\right) \right] + \left[ y^{\mathrm{GW}}_{1[13]3}(t_B) \right] \nonumber \\
&=& T_{13}\left( \left[ 1 + \hat{k}^{\mathrm{GW}}(t_B) \cdot \hat{n}_{[31]}(t_B) \right] \frac{1}{2} \cross \left\{ \Psi_{[31]} \left[ t_3(t_B) - \hat{k_B} \cdot \vec{p}_3(t_3(t_B))\right] - \Psi_{[31]} \left[ t_B - \hat{k}^{\mathrm{GW}}(t_B) \cdot \vec{p}_1(t_B) \right]  \right\} \right) \nonumber \\
&+& \left[ 1 + \hat{k}^{\mathrm{GW}}(t_B) \cdot \hat{n}_{[13]}(t_B) \right] \frac{1}{2} \cross \left\{ \Psi_{[13]} \left[ t_1(t_B) - \hat{k_B} \cdot \vec{p}_1(t_1(t_B))\right] - \Psi_{[13]} \left[ t_B - \hat{k}^{\mathrm{GW}}(t_B) \cdot \vec{p}_3(t_B) \right]  \right\} \label{eq:n1cH313}
\end{eqnarray}
\end{widetext}
where,
\begin{itemize}
    \item $y^{\mathrm{GW}}_{3[31]1}(t_B)$ is described below Eq.~\ref{eq:n1cH21}.
    \item $y^{\mathrm{GW}}_{1[13]3}(t_B)$ is the 2-pulse response time-series in a Doppler link from OB13 to OB31.
    \item $\Psi, \hat n, \hat k, \vec p_i, t_B$ are described under Eqs.~\ref{eq:dopplerlink},~\ref{eq:geomlink}.
    \item $t_1, t_3$ are time stamps from the two clocks in S/C1 and S/C3 respectively.
    \item $T_{13}$ is the physically occurring delay along the light beam travelling from optical bench OB13 to OB31.
\end{itemize}

Finally, for the remaining fourth beatnote $\mathcal{H}^{n\ell}_{323}$ can be understood similarly to the phasemeter measurement $\mathcal{H}^{n\ell}_{313}$ above. $\mathcal{H}^{n\ell}_{323}$, is the phasemeter measurement at OB32, whose laser beam paths are shown in \cref{fig:n1c_B323}. Concretely, this beatnote is formed by interfering (subtracting) a two-way beam along the path: PL32 $\rightarrow$ SL23 $\rightarrow$ SL32 with the local laser PL32 after a round-trip return. The doppler response of the binary signal is given by, with the clock reference of USO3, $t_3$:  
\begin{widetext}
\begin{eqnarray}
\mathcal{H}^{n\ell,\text{GW}}_{323}(t_3) &=& \left[ T_{23}\left( y^{\mathrm{GW}}_{3[32]2}(t_B) \right) \right] + \left[ y^{\mathrm{GW}}_{2[23]3}(t_B) \right]  \nonumber \\
&=& T_{23}\left( \left[ 1 + \hat{k}^{\mathrm{GW}}(t_B) \cdot \hat{n}_{[32]}(t_B) \right] \frac{1}{2} \cross \left\{ \Psi_{[32]} \left[ t_3(t_B) - \hat{k}^{\mathrm{GW}}(t_B) \cdot \vec{p}_3(t_3(t_B))\right] - \Psi_{[32]} \left[ t_B - \hat{k}^{\mathrm{GW}}(t_B) \cdot \vec{p}_2(t_B) \right] \right\} \right) \nonumber \\
&+& \left[ 1 + \hat{k}^{\mathrm{GW}}(t_B) \cdot \hat{n}_{[23]}(t_B) \right] \frac{1}{2} \cross \left\{ \Psi_{[23]} \left[ t_2(t_B) - \hat{k}^{\mathrm{GW}}(t_B) \cdot \vec{p}_2(t_2(t_B))\right] - \Psi_{[23]} \left[ t_B - \hat{k}^{\mathrm{GW}}(t_B) \cdot \vec{p}_3(t_B) \right] \right\} \label{eq:n1cB323dopplerlink}
\end{eqnarray}
\end{widetext}
where,
\begin{itemize}
    \item $y^{\mathrm{GW}}_{3[32]2}(t_B)$ is described below Eq.~\ref{eq:n1cH12}.
    \item $y^{\mathrm{GW}}_{2[23]3}(t_B)$ is the 2-pulse response time-series in a Doppler link from OB32 to OB23.
    \item $\Psi, \hat n, \hat k, \vec p_i, t_B$ are described under Eqs.~\ref{eq:dopplerlink},~\ref{eq:geomlink}.
    \item $t_2, t_3$ are time stamps from the two clocks in the S/C2 and S/C3 respectively.
    \item $T_{23}$ is the physically occurring delay along the light beam travelling from optical bench OB23 to OB32.
\end{itemize}

The results for the phasemeter equations are compared for two different locking schemes for V407Vul and ZTTFJ1539 below in Sec.~\ref{non_swap_max}. These GW coupling equations derived above will be used in Sec.~\ref{TDIvars} together with the exemplary GW parameters from Sec.~\ref{vbs} to compute the response of the time delay observables instead of the usual 2-pulse response for each arm obtained with free-running laser configuration for the case of simplified set of instrument noises. 
\subsection{Maximally locked non-swap configuration}\label{non_swap_max}
\begin{figure}[htb]
  \centering
    \includegraphics[width=0.5\textwidth]{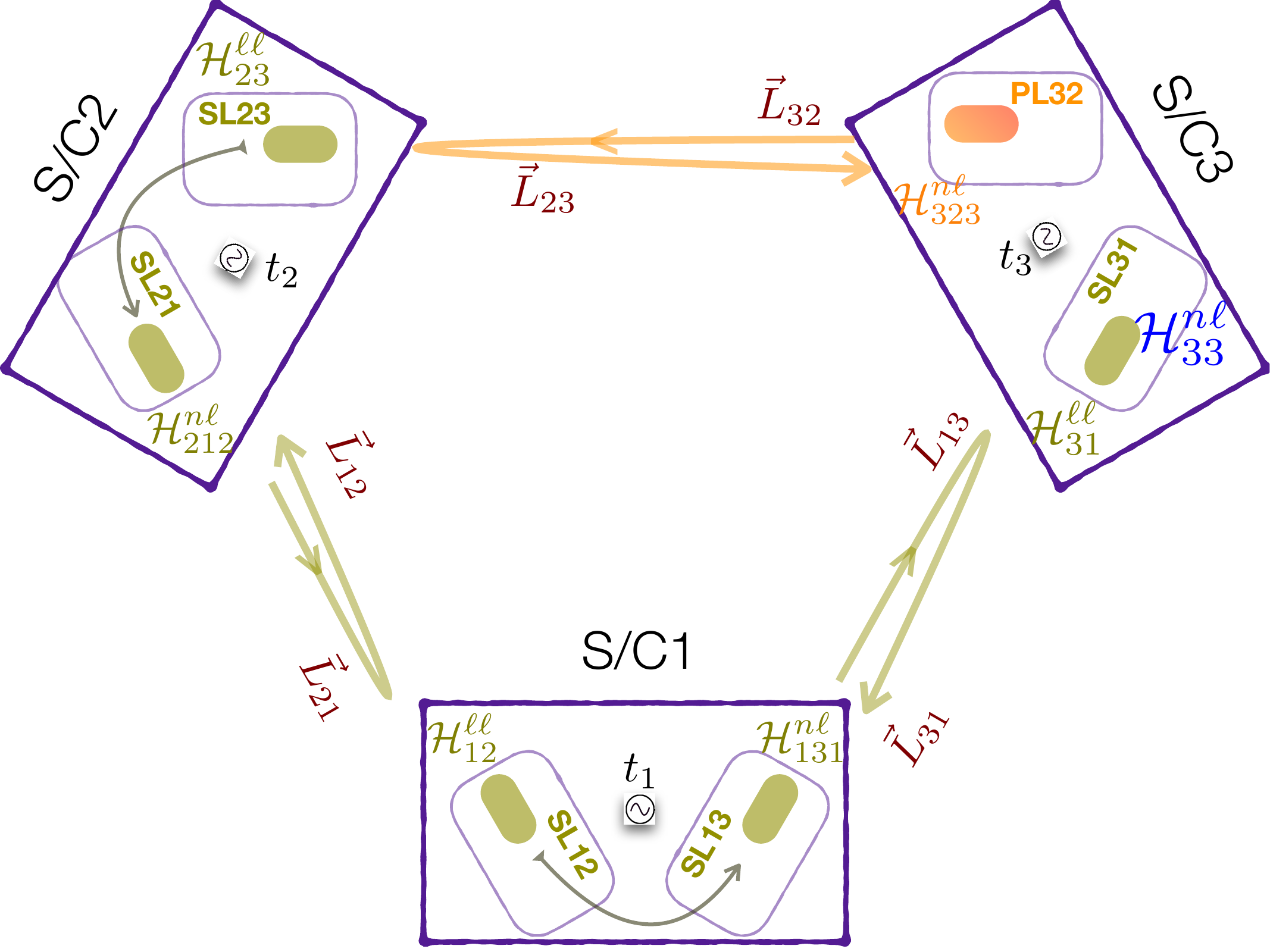}
  \caption{`N2a' Laser locking configuration with maximum number of Doppler shifts in the long-arm used to lock the primary laser, PL32 (in orange) in S/C3 in forming one of the four beatnotes, i.e. the maximum number of $D$ used to lock the secondary laser is three (adapted from \citep{Heinzel2020}). $\mathcal{H}^{\ell\ell}_{11}, \mathcal{H}^{\ell\ell}_{22},\mathcal{H}^{n\ell}_{33} $ reference beatnotes within S/C1, S/C2, S/C3 respectively. $\mathcal{H}^{\ell\ell}_{31}, \mathcal{H}^{\ell\ell}_{23}, \mathcal{H}^{n\ell}_{131}, \mathcal{H}^{n\ell}_{212}, \mathcal{H}^{n\ell}_{323}$ are the long-arm beatnotes across the constellation. Secondary lasers locked to PL32 are labelled by SL as in ~\cref{fig:n1c}. The grey arrows indicate the locking scheme for adjacent lasers within a S/C. $\vec L_{12}, \vec L_{21}, \vec L_{13}, \vec L_{31}, \vec L_{23}, \vec L_{32}$ are inter-S/C Doppler links or armlengths. \label{fig:n2a}}
\end{figure}

Similar to the configuration `N1c' (\cref{fig:n1c}) described in the subsection above, the resulting reference and long-arm beatnotes for the configuration `N2a' (\cref{fig:n2a}) are described here. We call this configuration `maximally locked' with reference to the beatnote signal, which has the maximum number of Doppler shifts in the long arm measurement in order to distinguish it from that of the `minimally locked', `N1c'. The figures below will make this clearer, where one of the beatnote signals that is formed in OB13 housed in S/C1 for the identical choice of primary laser as in the minimal locking configuration, has a total of eight instantaneous GW impinges as derived below in equations and demonstrated by laser beam paths (see, \cref{fig:n2a_H33}).

~\cref{fig:n2a} shows the topology of the beatnotes for the non-swap laser locking configuration, `N2a'. Unlike in \cref{fig:n1c}, this laser locking configuration has three, two and one Doppler links locked to PL32 (in orange) in S/C3 in forming three of the four long-arm beatnotes. The distinguishing feature of this configuration is that the one of the local reference/test-mass interferometric measurements, $\mathcal{H}^{n\ell}_{33}$ in S/C has the long-arm Doppler links which will contain the Verification Binary (VB) signal.

\begin{figure}[htb]
  \centering
    \includegraphics[width=0.5\textwidth]{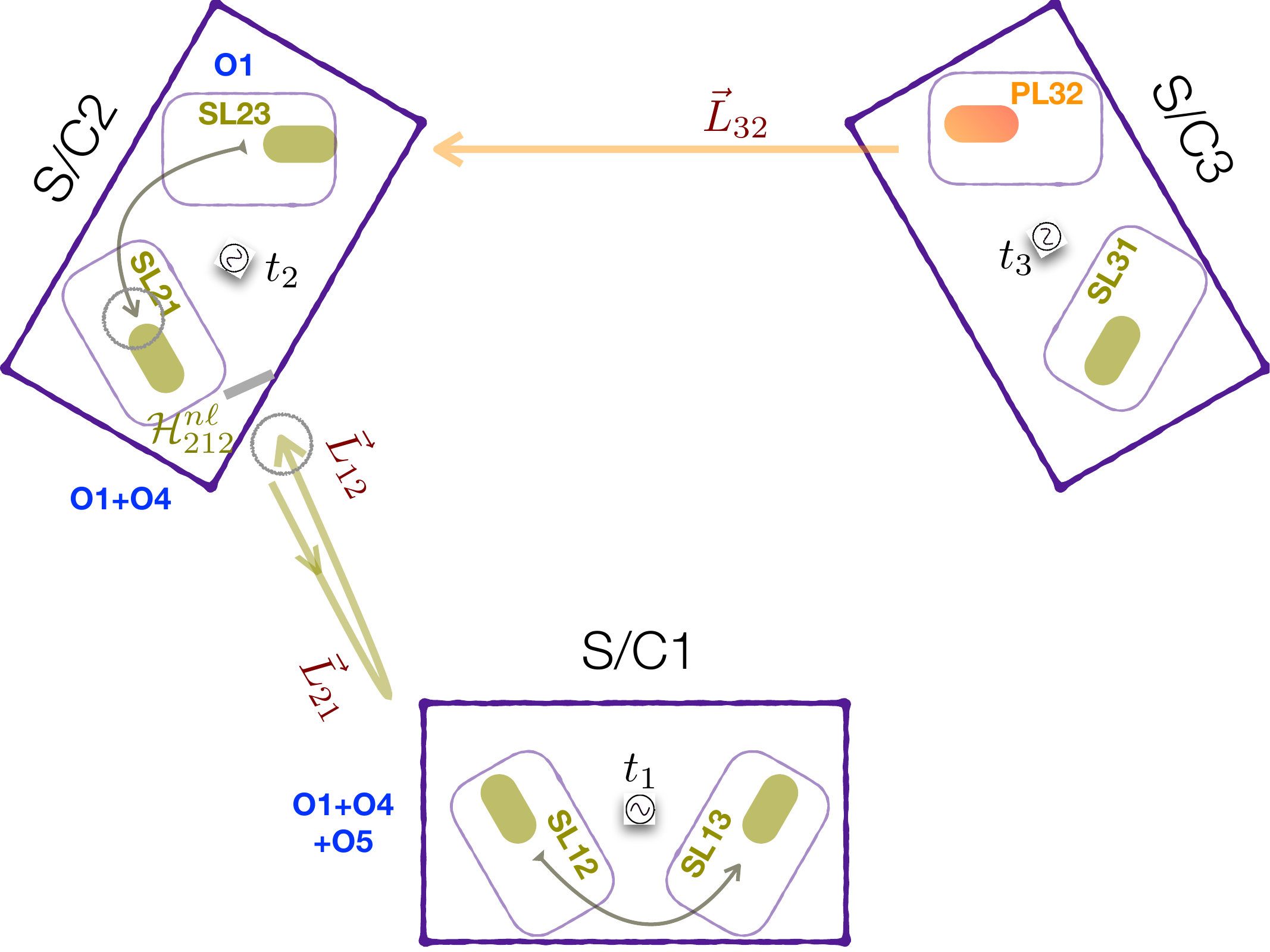}
  \caption{A pair of laser beams showing their respective paths along the long-arm (indicated by grey circles) in forming the interferometric measurement $\mathcal{H}^{n\ell}_{212}$ where the interference is indicated by slanted grey line at OB21. In this long arm interferometric measurement, there are two distinct beams subtracted from one another, however one of the beams has a two-way measurement as described in the text. \label{fig:n2a_H212}}
\end{figure}

As in the case of the minimal locking case, however, this configuration also has a total of four beatnote signals with the long arm which have non-zero GW signatures in them. Intriguingly the fourth beatnote signal with the long arm preserved non-zero VB GW signal in it is the local test-mass interferometer visually indicated in ~\cref{fig:n2a}, located in S/C3 by $\mathcal{H}^{n\ell}_{33}$. This has stirred a discussion within the community to abandon the nomenclature of the science interferometer to mean the long arm interferometer by default since the local interferometer without the explicit long arm Doppler shift can have science signal in it \citep{Heinzel2020}. The six long arm and reference/test mass interferometric measurements with the local offsets and laser frequency fluctuations are (adapted from \citep{Heinzel2020}):
    \begin{subequations}
    \label{eq:nswap2a}
        \begin{align}
        \mathcal{H}_{11} &= O2 \;\;, \mathcal{H}_{22} = O4\;\;,\label{eq:nswap2a:locrefs}\\
        \mathcal{H}^{\ell\ell}_{23} &= -O1\;\;, \mathcal{H}^{\ell\ell}_{31} = -O3 \;\; \mathcal{H}^{\ell\ell}_{12} = -O5,\label{eq:nswap2a:h23h31h12} \\
        \mathcal{H}^{n\ell}_{131} &= -f_{0,[21][32]} - f_{0,[31][13][21][32]} \nonumber \\
        &-D_{32,[21]} + D_{32,[31][13][21]} + D_{13,[31]} + D_{31} \nonumber \\
        & - D_{21} + D_{21,[31][13]} \nonumber \\
        &- O1_{,[21]} +O1_{,[31][13][21]}
        - O2 + O2_{,[31][13]}
        + O3_{,[31]}\nonumber \\
        &- O4_{,[21]} + O4_{,[31][13][21]}
        - O5 + O5_{[31][13]}, \label{eq:nswap2a:h131}\\
        \mathcal{H}^{n\ell}_{212} &= -f_{0,[32]} + f_{0,[12][21][32]}  \nonumber \\
        & - D_{32} + D_{32,[12][21]} + D_{13,[21]} + D_{12}  \nonumber \\
        &- O1 + O1_{[12][21]} - O4 + O4_{,[12][21]} + O5_{,[12]}, \label{eq:nswap2a:h212} \\
        \mathcal{H}^{n\ell}_{323} &= -f_0 + f_{0,[23][32]}  \nonumber \\
        &+ D_{32,[23]} +D_{23} \nonumber \\
        &+ O1_{,[23]}, \label{eq:nswap2a:h323} \\
        \mathcal{H}^{n\ell}_{33} &= f_0 - f_{0,[13][21][32]}  \nonumber \\
        &- D_{32,[13][21]} - D_{13} - D_{21,[13]} \nonumber \\
        &- O1_{,[13][21]} - O2_{,[13]} - O3 - O4_{,[13][21]} - O5_{,[13]}, \label{eq:nswap2a:h33}
        \end{align}
    \end{subequations}
where,
\begin{itemize}
  \item Eqs.~\ref{eq:nswap2a:locrefs} are the measurements of the local interferometers measuring the phase difference between two adjacent lasers on the two optical benches (OB).
  \item Eqs.~\ref{eq:nswap2a:h23h31h12} are the measurements of the long arm interferometers measuring the phase difference between two lasers on board two S/C without Doppler shifts in them, see \cref{fig:n2a_H212}.
  \item Eqs.~\ref{eq:nswap2a:h131}, ~\ref{eq:nswap2a:h212}, ~\ref{eq:nswap2a:h323}, are the measurements of the long arm interferometers measuring the phase difference between two lasers on board two S/C with varying number of  Doppler shifts in them, see \cref{fig:n2a_H212}, \cref{fig:n1c_B323}.
  \item Eq.~\ref{eq:nswap2a:h33} is the measurement of two local interferometers in S/C3 measuring the phase difference between its adjacent lasers that has non-vanishing Doppler links from the long arm, see \cref{fig:n2a_H33}.
  \item $f_{0}^{\ell}$ is laser frequency of the primary laser delayed by one link, measured according to clock in S/C1
  \item All two-index subscripts in $f_{0,[ij]}^{\ell}, O_{,[ij]}, D_{[ij]}$ are physically occurring delays along the armlength $\vec L_{ij}$.
  \item $D$ is Doppler shift at the time of arrival at the receiver.
  \item $O1 ... O5$ are sets of five distinct laser frequency offsets whose different linear combinations are applied to individual secondary lasers, see \cref{fig:n1c_B12}.
\end{itemize}
The rest of the lasers are secondary as shown in \cref{fig:n1c}. SL23 is locked to PL32 with a single Doppler shift. SL21 is locked to SL23 with offset O1. SL21 is locked to SL12 with offsets $O1+O4$. SL13 is locked to SL12 with offsets $O1+O4+O5$. SL31 is locked to SL13 four Doppler shifts and laser offsets $O1+O2+O3+O4+O5$. This last locking is the distinguishing feature of this configuration. There are three distinct long-arm measurements across the constellation shown by the inter-S/C laser links with arrows. Note, that unlike in the \cref{fig:n1c} in subsection Sec.~\ref{n1c_resp}, SL31 is not locked to PL32 despite being on the same S/C.
\subsection{Breakdown of individual beam paths for the maximally locked beatnotes}\label{n2a_resp}
\begin{figure}[htb]
  \centering
    \includegraphics[width=0.5\textwidth]{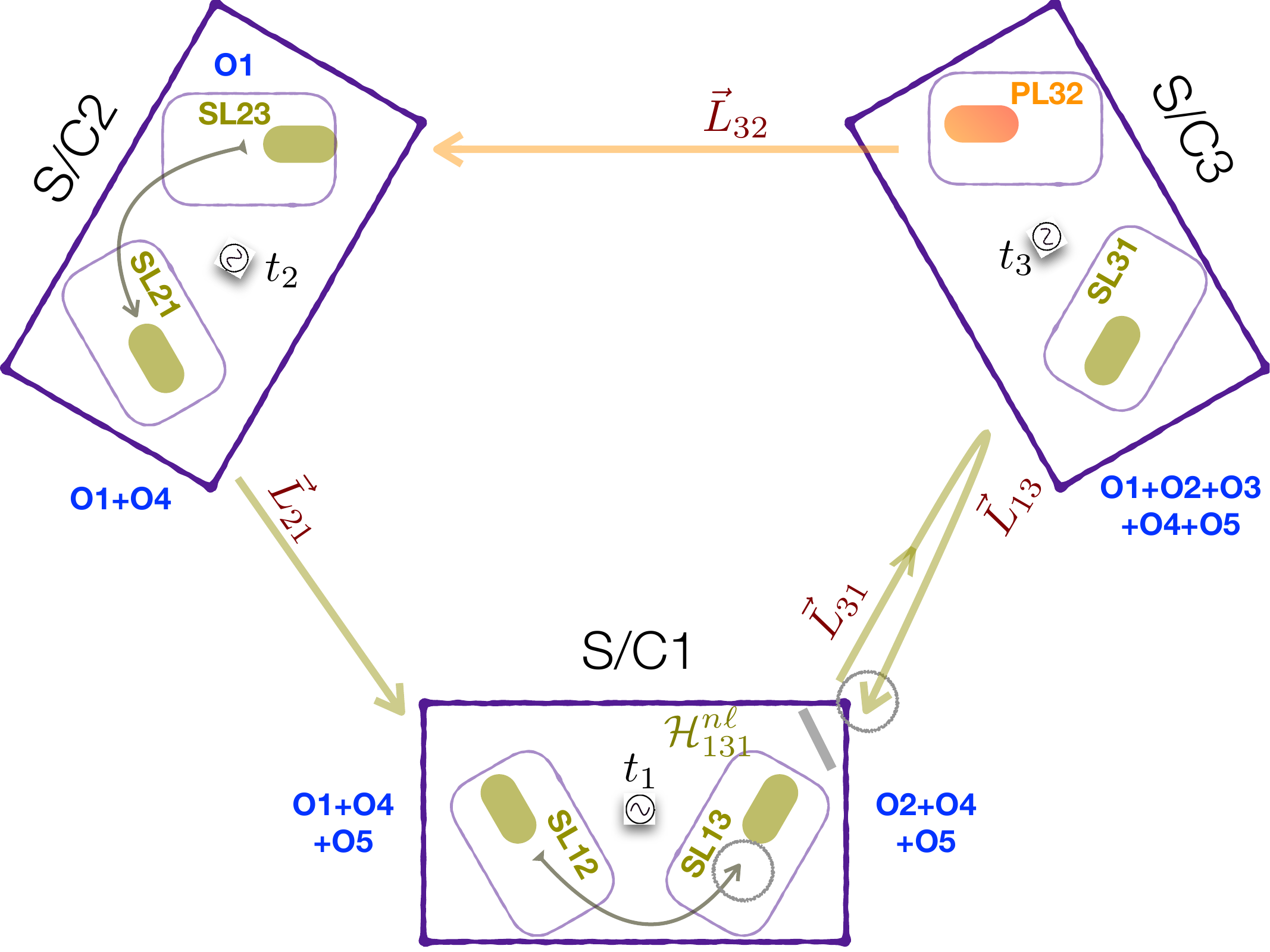}
  \caption{Similar to \cref{fig:n2a_H212}, where the long arm interference $\mathcal{H}^{n\ell}_{131}$ is measured on OB12 in S/C 1 using two distinct beam paths. One of the beams have two single arm Doppler links whereas the other beam has all three Doppler links including one two-way path.
\label{fig:n2a_H131}}
\end{figure}

\begin{figure}[htb]
  \centering
    \includegraphics[width=0.5\textwidth]{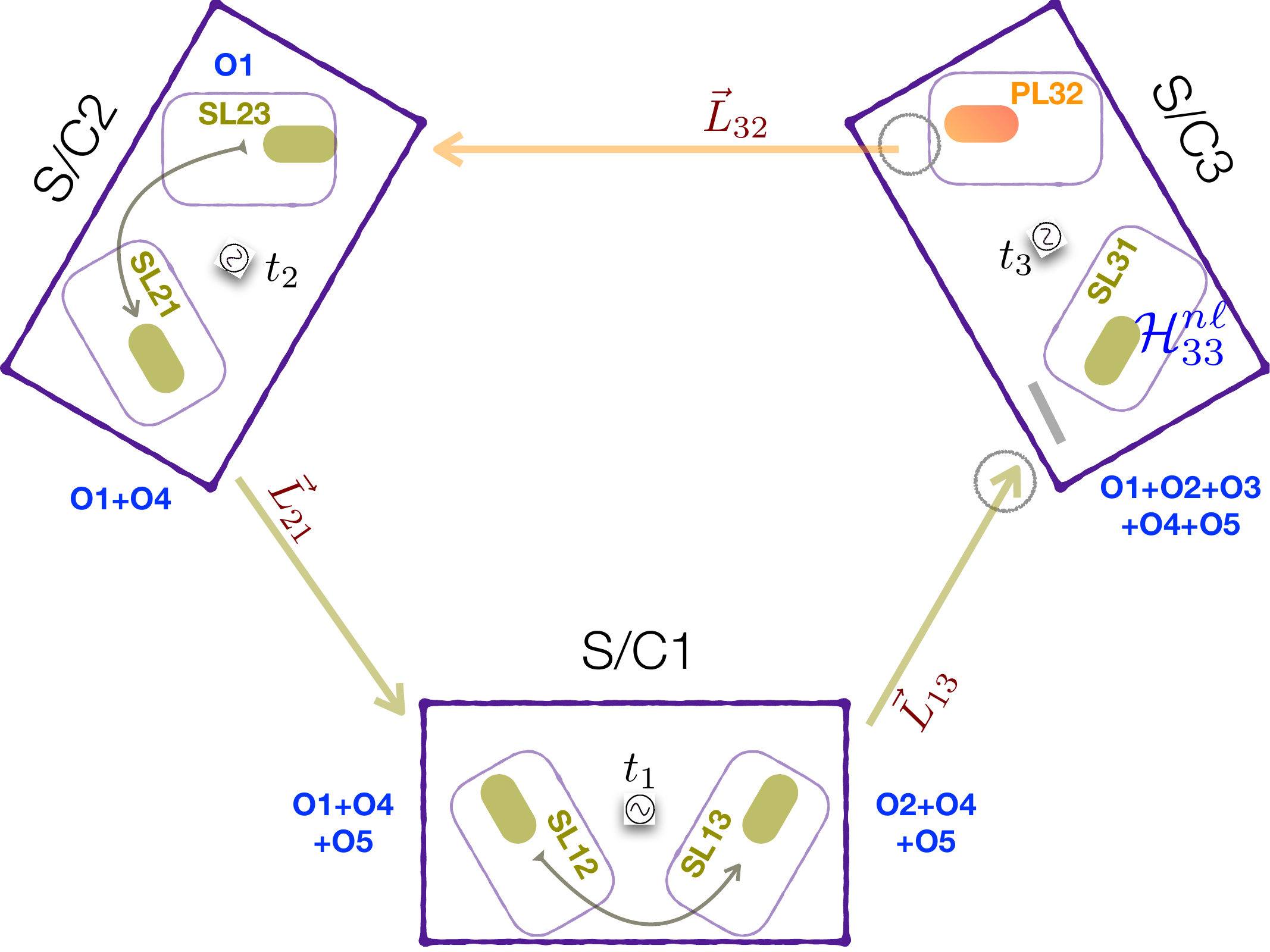}
  \caption{Unlike in \cref{fig:n2a_H212}, and \cref{fig:n2a_H131}, this is a reference (or test mass) interference $\mathcal{H}^{n\ell}_{33}$ measured on OB31 in S/C 3. It is also constructed using two distinct beam paths with single paths for both beams as described in the text.
\label{fig:n2a_H33}}
\end{figure}

Similar to Sec.~\ref{n1c_resp} above, there are four distinct long arm and local interferometric measurements  for the laser offset locking configuration `N2a', shown in \cref{fig:n2a} with varying number of Doppler shifts in any one of them. One of them, $\mathcal{H}^{n\ell}_{323}(t_3)$ (Eq.~\ref{eq:nswap2a:h323}, \cref{fig:n1c_B323}) in subsection above is identical to Eqs.~\ref{eq:nswap1c:h323},~\ref{eq:n1cB323dopplerlink}, thus making this beatnote common to both the minimally locked (`N1c') and maximally locked (`N2a') configurations. The remaining three beatnotes for which the beam paths are different with respect to `N1c' configuration are shown in \cref{fig:n2a_H212}, \cref{fig:n2a_H131}, and \cref{fig:n2a_H33}. According to the non-swap \texttt{fplan} derived in \citep{Heinzel2020}, the beatnote, $\mathcal{H}^{n\ell}_{323}(t_3)$ is common to five of the six configurations, except one of the laser offset configurations, named `N1b' (to be studied in a follow up paper). 

Thus, in constructing the TDI observables in the subsequent section, we can use the GW induced response to the phasemeter output from Eq.~\ref{eq:n1cB323dopplerlink} without any change for the corresponding solutions for configuration `N2a'.
%
A comparative version of Eq.~\ref{eq:nswap_h12_beam_paths} for the laser beam paths of the interferometric measurement of  $\mathcal{H}^{n\ell}_{212}$ in Eq.~\ref{eq:nswap2a:h212} is derived below and the laser beams path used to construct that interference is shown in \cref{fig:n2a_H212}. This beatnote on OB21 is formed by interfering one single path beam: PL32 $\rightarrow$ SL23 $\rightarrow$ SL21  and another single path beam which has a roundtrip beam: PL32 $\rightarrow$ SL23 $\rightarrow$ SL21 $\rightarrow$ SL12 $\rightarrow$ SL21, given by:
\begin{eqnarray}
\mathcal{H}^{n\ell}_{212} &=& [T_{21}(\text{SL}12)] \;\;\;\;\;\;\;\;\;\;\;\;\;\;\;\;\;\;\;\;\; - [\text{SL}21] \nonumber \\ 
&=& T_{32}[T_{21}[T_{12}(\text{PL}32)]] \;\;\;\;\;\;\; - T_{32}[\text{PL}32] \nonumber \\ 
&=& T_{32}[T_{21}[T_{12}(f_{0}^{\ell})]] \;\;\;\;\;\;\;\;\;\;\;\; - T_{31}(f_{0}^{\ell})  \nonumber \\ 
&+& T_{12}[T_{21}[O_1]] - O_4 + T_{12}[T_{21}[O_4]] + T_{12}[O_5] - O_1 \nonumber \\
&-&D_{32}^{\text{VB}} + T_{12}[T_{21}(D_{32}^{\text{VB}})] + T_{12}[D_{21}^{\text{VB}}] + D_{12}^{\text{VB}},  \label{eq:nswap_h212_n2a_beam_paths}
\end{eqnarray}
where,
\begin{itemize}
  \item $\mathcal{H}^{n\ell}_{212}$ is the long arm beatnote measured on OB21, where distant beam from laser on OB12 is subtracted from the local laser beam OB21.
  \item $f_{0}^{\ell}, O1, ... O5$ are described below Eqs.~\ref{eq:nswap1c},~\ref{eq:nswap_h12_VBdoppler}.
  \item $D_{ij}^{\text{VB}}$ is Doppler shift of the VB signal at the time of arrival at the receiver, $j$.
  \item $T_{ij}$ is delay operator for light travel time along the arm travelling from $i \rightarrow j$, note that this is written alternatively in form of $[ij]$ as subscript, for eg. in Eqs.\ref{eq:nswap1c:h12}, \ref{eq:nswap1c:h21}, \ref{eq:nswap1c:h313}, \ref{eq:nswap1c:h323}, \ref{eq:nswap_h12_VBdoppler} and elsewhere.
\end{itemize}
The beam paths for interference for $\mathcal{H}^{n\ell}_{131}$ in the LISA constellation is shown figuratively in \cref{fig:n2a_H131} below. Here, this  beatnote at OB13 is formed by a interfering a one-way and path beam of: PL32 $\rightarrow$ SL23 $\rightarrow$ SL21 with the two-way path beam of: PL32 $\rightarrow$ SL31 $\rightarrow$ SL13 $\rightarrow$ SL31.
Substituting only the contributions from GW using Eqs.\ref{eq:dopplerlink}, \ref{eq:geomlink}, and with the clock reference of USO2, $t_2$, we get:
\begin{widetext}
\begin{align}
\mathcal{H}^{n\ell,\text{GW}}_{212}(t_2) =  - y^{\text{GW}}_{3[32]2}(t_B) + T_{12}\left(T_{21}\left( y^{\text{GW}}_{3[32]2}(t_B)\right) \right) + T_{12}\left( y^{\text{GW}}_{2[21]1}(t_B)\right) + y^{\text{GW}}_{1[12]2}(t_B) \nonumber \\
= - \left[ 1 + \hat{k}^{\text{GW}}(t_B) \cdot \hat{n}_{[32]}(t_B) \right]\frac{1}{2} \cross \left\{ \Psi_{[32]} \left[ t_3(t_B) - \hat{k}^{\text{GW}}(t_B) \cdot \vec{p}_3(t_3(t_B))\right] - \Psi_{[32]} \left[ t_B - \hat{k}^{\text{GW}}(t_B) \cdot \vec{p}_2(t_B) \right]  \right\} \nonumber \\
+ T_{12}\left( T_{21} \left( \left[ 1 + \hat{k}^{\text{GW}}(t_B) \cdot \hat{n}_{[32]}(t_B) \right]\frac{1}{2} \cross \left\{ \Psi_{[32]} \left[ t_3(t_B) - \hat{k}^{\text{GW}}(t_B) \cdot \vec{p}_3(t_3(t_B))\right] - \Psi_{[32]} \left[ t_B - \hat{k}^{\text{GW}}(t_B) \cdot \vec{p}_2(t_B) \right]  \right\} \right) \right) \nonumber \\
+ \left[ 1 + \hat{k}^{\text{GW}}(t_B) \cdot \hat{n}_{[21]}(t) \right] \frac{1}{2} \cross \left\{ \Psi_{[21]} \left[ t_2(t_B) - \hat{k}^{\text{GW}}(t_B) \cdot \vec{p}_2(t_2(t_B))\right] - \Psi_{[21]} \left[ t_B - \hat{k}^{\text{GW}}(t_B) \cdot \vec{p}_1(t_B) \right]  \right\} \nonumber \\
+ \left[ 1 + \hat{k}^{\text{GW}}(t_B) \cdot \hat{n}_{[12]}(t) \right] \frac{1}{2} \cross \left\{ \Psi_{[12]} \left[ t_1(t_B) - \hat{k}^{\text{GW}}(t_B) \cdot \vec{p}_1(t_1(t_B))\right] - \Psi_{[12]} \left[ t_B - \hat{k} \cdot \vec{p}_2(t_B) \right]  \right\} \label{eq:n2aB212dopplerlink}
\end{align}
\end{widetext}
where,
\begin{itemize}
    \item $y^{\text{GW}}_{3[32]2}(t_B)$ is the 2-pulse response time-series in a Doppler link from OB32 to OB23.
    \item $y^{\mathrm{GW}}_{2[21]1}(t_B)$ is the 2-pulse response time-series in a Doppler link from OB21 to OB12.
    \item $y^{\mathrm{GW}}_{1[12]2}(t_B)$ is the 2-pulse response time-series in a Doppler link from OB12 to OB21.
    \item $\Psi, \hat n, \hat k, \vec p_i, t_B$ are described under Eqs.~\ref{eq:dopplerlink},~\ref{eq:geomlink}.
    \item $t_1, t_2, t_3$ are time stamps from the three clocks in the S/C1, S/C2, and S/C3 respectively.
    \item $T_{12}$ and $T_{21}$ are the physically occurring delays along the light beam travelling from optical bench OB12 to OB21 and the same for OB21 to OB12 respectively.
\end{itemize}
%
Similar to the GW induced only measurement $\mathcal{H}^{n\ell}_{212}$ above, the beam paths for interference for $\mathcal{H}^{n\ell}_{131}$ in the LISA constellation is shown figuratively in \cref{fig:n2a_H131} below (Eq.~\ref{eq:nswap2a:h131}). Here, this  beatnote is measured on OB21 which is formed by a interfering a beam with the path of: PL32 $\rightarrow$ SL23 $\rightarrow$ SL21 $\rightarrow$ SL12 $\rightarrow$ SL13 with the beam path consisting of a two-way path of: PL32 $\rightarrow$ SL23 $\rightarrow$ SL21 $\rightarrow$ SL12 $\rightarrow$ SL13 $\rightarrow$ SL31 $\rightarrow$ SL13. The corresponding Doppler response of the binary signal, with the clock reference of USO1, $t_1$, is given by:
\begin{widetext}
\begin{align}
\mathcal{H}^{n\ell,\text{GW}}_{131}(t_1) = - T_{21}\left(y^{\mathrm{GW}}_{3[32]2}(t_B)\right) + T_{31}\left(T_{13}\left(T_{31}\left(y^{\mathrm{GW}}_{3[32]2}(t_B)\right)\right)\right) + T_{31}\left( y^{\mathrm{GW}}_{1[13]3}(t_B)\right) + y^{\mathrm{GW}}_{3[31]1}(t_B)  \nonumber \\
- y^{\mathrm{GW}}_{2[21]1}(t_B) + T_{31}\left( T_{13}\left(y^{\mathrm{GW}}_{2[21]1}(t_B)\right) \right) \nonumber \\
= - T_{21}\left(\left[ 1 + \hat{k}^{\text{GW}}(t_B) \cdot \hat{n}^{\ell}_{32}(t_B) \right] \cross \frac{1}{2}\left\{ \Psi_{32} \left[ t_{3}(t_B) - \hat{k}^{\text{GW}}(t_B) \cdot \vec{p}_{3}(t_{3}(t_B)) \right] - \Psi_{32} \left[ t_B - \hat{k}^{\text{GW}}(t_B) \cdot \vec{p}_{2}(t_B) \right]  \right\} \right) \nonumber \\
+ T_{31}\left(T_{13}\left(T_{31}\left(\left[ 1 + \hat{k}^{\text{GW}}(t_B) \cdot \hat{n}^{\ell}_{32}(t_B) \right] \cross \frac{1}{2}\left\{ \Psi_{32} \left[ t_{3}(t_B) - \hat{k}^{\text{GW}}(t_B) \cdot \vec{p}_{3}(t_{3}(t_B)) \right] - \Psi_{32} \left[ t_B - \hat{k}^{\text{GW}}(t_B) \cdot \vec{p}_{2}(t_B) \right]  \right\} \right)\right)\right)\nonumber \\
+ T_{31}\left(\left[ 1 + \hat{k}^{\text{GW}}(t_B) \cdot \hat{n}^{\ell}_{13}(t_B) \right] \cross \frac{1}{2}\left\{ \Psi_{13} \left[ t_{1}(t_B) - \hat{k}^{\text{GW}}(t_B) \cdot \vec{p}_{1}(t_1(t_B)) \right] - \Psi_{13} \left[ t_B - \hat{k}^{\text{GW}}(t_B) \cdot \vec{p}_{3}(t_B) \right]  \right\}\right) \nonumber \\
+ \left[ 1 + \hat{k}^{\text{GW}}(t_B) \cdot \hat{n}^{\ell}_{31}(t_B) \right] \cross \frac{1}{2}\left\{ \Psi_{31} \left[ t_{\text{s}}(t_B) - \hat{k}^{\text{GW}}(t_B) \cdot \vec{p}_{3}(t_{3}(t_B)) \right] - \Psi_{31} \left[ t_B - \hat{k}^{\text{GW}}(t_B) \cdot \vec{p}_{1}(t_B) \right]  \right\} \nonumber \\
- \left[ 1 + \hat{k}^{\text{GW}}(t_B) \cdot \hat{n}^{\ell}_{21}(t_B) \right] \cross \frac{1}{2}\left\{ \Psi_{21} \left[ t_{\text{s}}(t_B) - \hat{k}^{\text{GW}}(t_B) \cdot \vec{p}_{\text{2}}(t_{\text{2}}(t_B)) \right] - \Psi_{21} \left[ t_B - \hat{k}^{\text{GW}}(t_B) \cdot \vec{p}_{\text{1}}(t_B) \right]  \right\} \nonumber \\
+ T_{13}\left( T_{31} \left( \left[ 1 + \hat{k}^{\text{GW}}(t_B) \cdot \hat{n}^{\ell}_{21}(t_B) \right] \cross \frac{1}{2}\left\{ \Psi_{21} \left[ t_{\text{s}}(t_B) - \hat{k}^{\text{GW}}(t_B) \cdot \vec{p}_{\text{2}}(t_{\text{2}}(t_B)) \right] - \Psi_{21} \left[ t_B - \hat{k}^{\text{GW}}(t_B) \cdot \vec{p}_{\text{1}}(t_B) \right]  \right\} \right) \right) \label{eq:n2aB131dopplerlink}
\end{align}
\end{widetext}
where,
\begin{itemize}
    \item $y^{\mathrm{GW}}_{3[32]2}(t_B), y^{\mathrm{GW}}_{2[21]1}(t_B)$ are described below Eq.~\ref{eq:n2aB212dopplerlink}.
    \item $y^{\mathrm{GW}}_{1[13]3}(t_B)$ is the 2-pulse response time-series in a Doppler link from OB13 to OB31.
    \item $y^{\mathrm{GW}}_{3[31]1}(t_B)$ is the 2-pulse response time-series in a Doppler link from OB31 to OB13.
    \item $\Psi, \hat n, \hat k, \vec p_i, t_B$ are described under Eqs.~\ref{eq:dopplerlink},~\ref{eq:geomlink}.
    \item $t_1, t_2, t_3$ are time stamps from the three clocks in the S/C1, S/C2, and S/C3 respectively.
    \item $T_{31}, T_{13}, T_{21}$ are the physically occurring delays along the light beam travelling from optical bench OB31 to OB13, OB13 to OB31, and the same for OB21 to OB12 respectively.
\end{itemize}

Finally, the fourth remaining beatnote frequency for configuration `N2a' is the local beatnote in OB33, namely, $\mathcal{H}^{n\ell}_{33}$. The beam paths for interference for this measurement in the LISA constellation is shown figuratively in \cref{fig:n2a_H33} below (Eq.~\ref{eq:nswap2a:h33}), where this beatnote is measured on by interfering a one-way beam with the path of: PL32 $\rightarrow$ SL23 $\rightarrow$ SL21 $\rightarrow$ SL12 $\rightarrow$ SL13 with the beam path consisting of another one-way beam of the path of: PL32 $\rightarrow$ SL23. The corresponding Doppler response of the binary signal, with the clock reference of USO3, $t_3$, is given by:
\begin{widetext}
\begin{align}
\mathcal{H}^{n\ell,\text{GW}}_{33}(t_3) = - T_{13}\left(T_{21}\left(y^{\mathrm{GW}}_{3[32]2}(t_B)\right)\right) -  y^{\mathrm{GW}}_{1[13]3}(t_B) - T_{13}\left( y^{\mathrm{GW}}_{2[21]1}(t_B)\right) \nonumber \\
= - T_{13}\left(T_{21}\left(\left[ 1 + \hat{k}^{\text{GW}}(t_B) \cdot \hat{n}^{\ell}_{32}(t_B) \right] \cross \frac{1}{2}\left\{ \Psi_{32} \left[ t_{3}(t_B) - \hat{k}^{\text{GW}}(t_B) \cdot \vec{p}_{3}(t_{3}(t_B)) \right] - \Psi_{32} \left[ t_B - \hat{k}^{\text{GW}}(t_B) \cdot \vec{p}_{2}(t_B) \right]  \right\} \right)\right) \nonumber \\
- \left[ 1 + \hat{k}^{\text{GW}}(t_B) \cdot \hat{n}^{\ell}_{13}(t_B) \right] \cross \frac{1}{2}\left\{ \Psi_{13} \left[ t_{1}(t_B) - \hat{k}^{\text{GW}}(t_B) \cdot \vec{p}_{1}(t_1(t_B)) \right] - \Psi_{13} \left[ t_B - \hat{k}^{\text{GW}}(t_B) \cdot \vec{p}_{3}(t_B) \right]  \right\} \nonumber \\
- T_{13}\left(\left[ 1 + \hat{k}^{\text{GW}}(t_B) \cdot \hat{n}^{\ell}_{21}(t_B) \right] \cross \frac{1}{2}\left\{ \Psi_{21} \left[ t_{\text{s}}(t_B) - \hat{k}^{\text{GW}}(t_B) \cdot \vec{p}_{\text{2}}(t_{\text{2}}(t_B)) \right] - \Psi_{21} \left[ t_B - \hat{k}^{\text{GW}}(t_B) \cdot \vec{p}_{\text{1}}(t_B) \right]  \right\}\right) 
\label{eq:n2aB33dopplerlink}
\end{align}
\end{widetext}
where,
\begin{itemize}
    \item $y^{\mathrm{GW}}_{3[32]2}(t_B), y^{\mathrm{GW}}_{1[13]2}(t_B), y^{\mathrm{GW}}_{2[21]1}(t_B)$ are described below Eq.~\ref{eq:n2aB212dopplerlink}.
    \item $\Psi, \hat n, \hat k, \vec p_i, t_B$ are described under Eqs.~\ref{eq:dopplerlink},~\ref{eq:geomlink}.
    \item $t_1, t_2, t_3$ are time stamps from the three clocks in the S/C1, S/C2, and S/C3 respectively.
    \item $T_{31}, T_{21}$ are the physically occurring delays along the light beam travelling from optical bench OB31 to OB13, and the same for OB21 to OB12 respectively.
\end{itemize}
\begin{figure*}
    \begin{minipage}[!t]{\textwidth}
        \includegraphics[width=\textwidth]{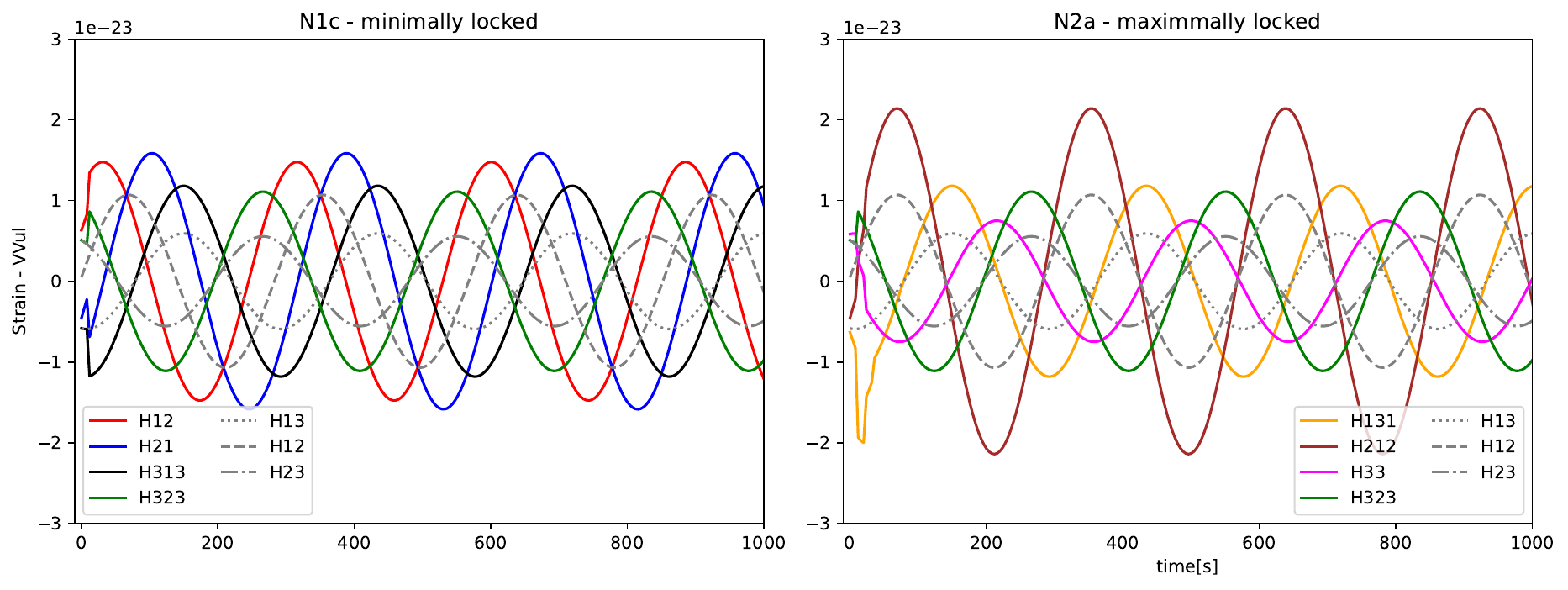}
\caption{Phasemeter output for long arm measurements for V407Vul. Left: non-locking long arm measurements for the N1c frequency plan. Right: non-locking long arm measurements for the N2a frequency plan. The comparison against long-arm measurements obtained using free running laser configuration along the three distinct arms in dashed-grey curves as labelled. The EM parameters for the VBs are taken from Table~\ref{tab:VBsparam}, assumed perfectly known for the simulations above. The simulation is shown for $\sim 16$min with sampling time of $dt=4$s.}
\label{fig:VVul_PM}
\end{minipage}
    \hfill
\end{figure*}
\begin{figure*}
    \begin{minipage}[!b]{\textwidth}
        \includegraphics[width=\textwidth]{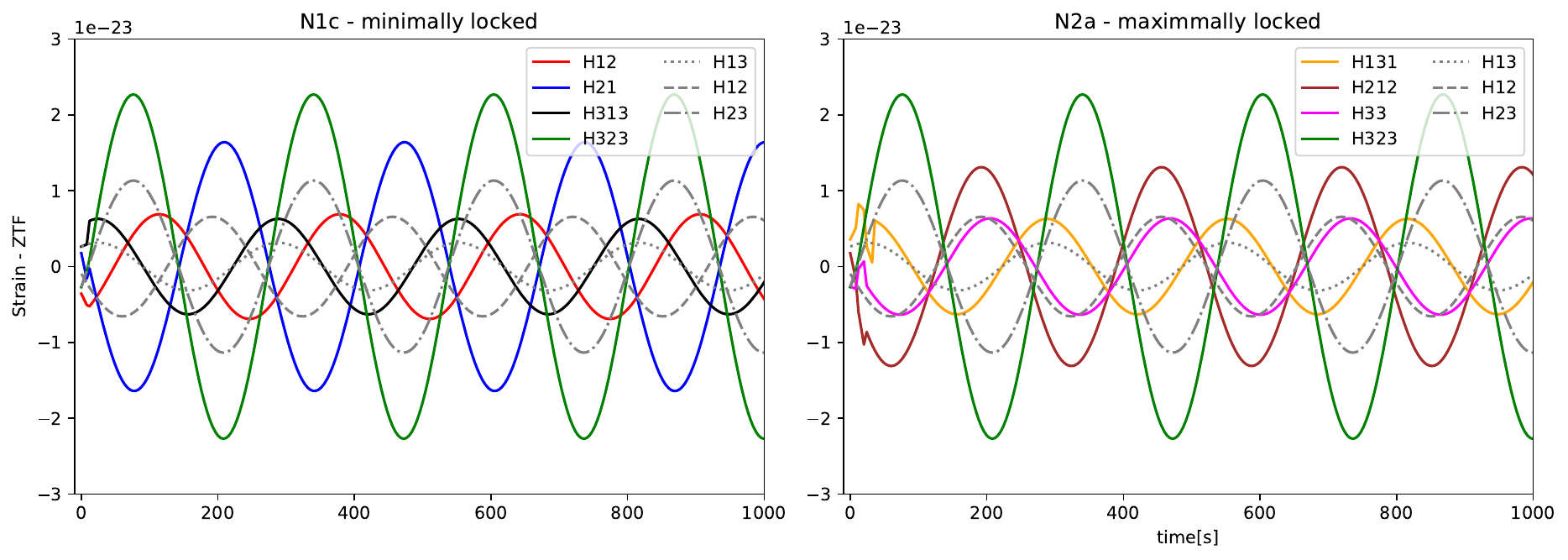}
\caption{Similar to \cref{fig:VVul_PM} above, phasemeter output from long arm measurements for ZTFJ2243. Left: non-locking long arm measurements for the N1c frequency plan. Right: non-locking long arm measurements for the N2a frequency plan. The comparison against long-arm measurements obtained using free running laser configuration along the three distinct arms in dashed-grey curves as labelled. }
\label{fig:ZTF_PM} 
\end{minipage}
    \hfill
\end{figure*}
These results for the phasemeter output above are shown in \cref{fig:VVul_PM} and \cref{fig:ZTF_PM} for the examples of face-on binary, V407Vul, and the eclipser, ZTTFJ1539 respectively. The signal coupling the non-locking long arm measurements are shown in solid-coloured curves for the minimally locked configuration, `N1c' on the left and for the maximally locked configuration, `N2a' on the left. Comparison with the response to six-one way links with free-running laser configurations are shown in grey coloured curves. Observe that the four non-locking phasemeter time-series for the two different locking schemes considered are drastically different than those using the free-running lasers. 

Overall the amplitude of the eclipsing system is higher as expected since its GW strain amplitude is larger than that for VVul by a factor of $\sim 1.3$. For VVul the higher level in the amplitudes for the `N2a' configuration is obvious which is attributable to the larger number of long-arm Doppler locks in their phasemeter measurements. The eclipsing system has comparable levels of the amplitude coupling for both configurations.

The two-way measurement, $\mathcal{H}^{n\ell,\text{GW}}_{323}$ is identical for both configurations and thus their response to each of the binaries are identical as shown by the  green curves. From these simple time-series analysis, one conclusion is that for `N2a' V407Vul, the measurement $\mathcal{H}^{n\ell,\text{GW}}_{212}$ can be utilised to calibrate the instrument due to its strongest amplitude. For the ZTTFJ1539 source, $\mathcal{H}^{n\ell,\text{GW}}_{323}$ is the strongest amplitude for both configurations and thus, this measurement (\cref{fig:ZTF_PM}) can be used for instrument calibration. However, since all of these signals will be buried in 10 orders of magnitude of laser noise, we need TDI signals, described in the following. 

As mentioned above, we will use the GW coupling to the phasemeter equations above to propagate them through Sec.~\ref{TDIvars} for the given set of two white dwarf binaries (Sec.~\ref{vbs}) to compute their response in the TDI variables in the subsection below.
\subsection{TDI observables}\label{TDIvars}
In this section we analyse the coupling of the signals from Secs.~\ref{non_swap_min}-~\ref{n2a_resp}, with respect to laser frequency fluctuations into the well known Time Delay Interferometry (TDI) Michelson type variable. We consider constant but unequal arms and use the long arm non-locking measurements for the two locking configurations described above as `minimally locked' and `maximally locked'. The exploration of a full set of TDI observables as a consequence of locking that would suppress the primary noises to the level of secondary noises is outside the scope of this study which will be studied in forthcoming work. For free-running laser configuration, a most recent extensive version of TDI variables can be found in \citep{Hartwig2022}. 
\begin{figure}[htb]
  \centering
    \includegraphics[width=0.45\textwidth]{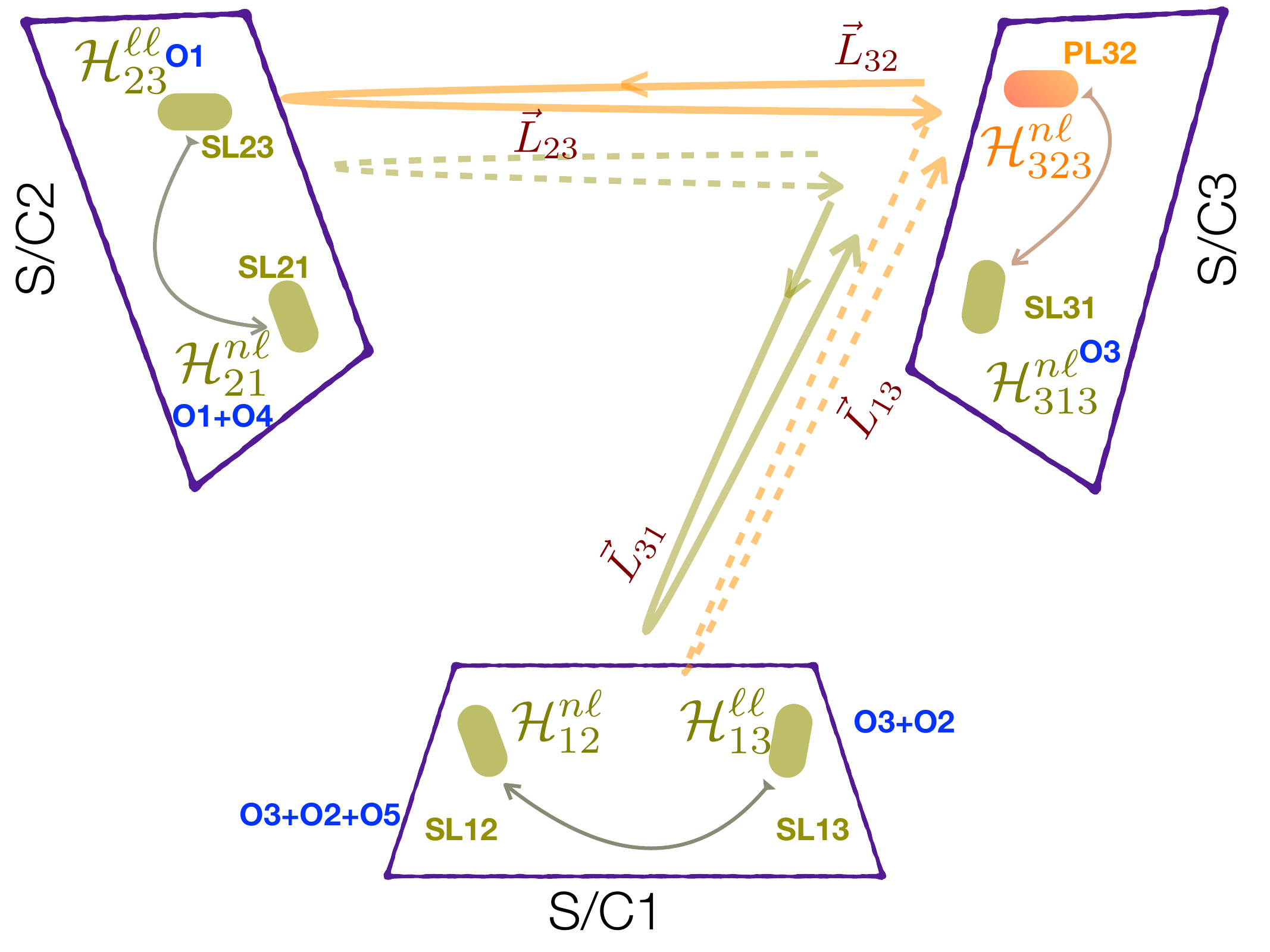}
  \caption{TDI X for N1c recovered using phasemeter measurements in solid curves. The delays applied to the measurements are shown in dashed lines.
\label{fig:n1c_X}}
\end{figure}
\subsection{N1c}
In order to compare the two configurations, one needs to compute SNR for TDI observable(s) that are common to both of them. A conundrum arises immediately - with the geometrically non-identical and non-overlapping measurements between the two topologies - what are the TDI observables common to both configurations? This question is out of scope for this study, to be explored in forthcoming work. Nonetheless, in this study we construct a pair of two distinct TDI variables for each topology that are comparable with regard to the total number of armlengths involved such that the differences in the TDI observables due to signal will be driving the main difference instead of the laser frequency fluctuations.

In order to verify the similarity or the difference between a given TDI observable that is common to both configurations one option is to answer the following - What is an equal pathlength Michelson (X, Y, Z) data stream for given measurements for `N1c'? For the minimally locked scheme, the classical TDI Z from the free-running laser configuration can be constructed with the two 2-way measurements to form a virtual Michelson interferometer centered at S/C 3. This can be constructed by subtracting $\mathcal{H}^{n\ell}_{323}$ delayed by $T_{31} T_{13}$ from $\mathcal{H}^{n\ell}_{313}$ delayed by $T_{32} T_{23}$, in the following:
\begin{widetext}
\begin{align}
    \label{eq:nswap1c_X}
    Z(t_B)^{\text{N1c}} &= \mathcal{H}^{n\ell}_{323, [31][13]} - \mathcal{H}^{n\ell}_{313, [32][23]} \nonumber \\
    &=-f_{0, [31][13]} + f_{0,[23][32][31][13]} +f_{0, [32][23]} - f_{0,[31][13][32][23]} \nonumber \\
    &+D_{32,[23][31][13]} + D_{23, [31][13]} - D_{13, [32][23]} - D_{31,[13][32][23]} \nonumber \\
    & + O1_{,[23], [32][23]} + O2_{, [13][32][23]} + O3_{, [32][23]} - O3_{,[31][13][32][23]}.
\end{align}
\end{widetext}
The local and long long-arm locking measurements in Eqs.~\ref{eq:nswap1c:locrefs},~\ref{eq:nswap1c:h23h13}, can be used to subtract the offset terms to get residual laser frequency noise and differential VB signal in the Eq.~\ref{eq:nswap1c_X} above, such that:  
\begin{widetext}
\begin{align}
    \label{eq:nswap1c_X_off}
    Z(t_B)^{\text{N1c}} &=-f_{0, [31][13]} + f_{0,[23][32][31][13]} +f_{0, [32][23]} - f_{0,[31][13][32][23]} \nonumber \\
    &+D_{32,[23][31][13]} + D_{23, [31][13]} - D_{13, [32][23]} - D_{31,[13][32][23]} \nonumber \\
    & + O1_{,[23], [32][23]} + O2_{, [13][32][23]} + O3_{, [32][23]} - O3_{,[31][13][32][23]}  \nonumber \\
    &+ \mathcal{H}^{\ell\ell}_{23, [32][23]}+ \mathcal{H}^{\ell\ell}_{13, [32][23]} + \mathcal{H}^{\ell\ell}_{33, [32][23]} - \mathcal{H}^{\ell\ell}_{33, [31][13][32][23]} \nonumber \\
    &= \mathcal{H}^{n\ell,\text{GW}}_{323, [31][13]} - \mathcal{H}^{n\ell,\text{GW}}_{313, [32][23]} -f_{0, [31][13]} + f_{0,[23][32][31][13]} +f_{0, [32][23]} - f_{0,[31][13][32][23]} \nonumber \\
    & =  \left[ T_{23}\left( y^{\mathrm{GW}}_{3[32]2}(t_B) \right) +  y^{\mathrm{GW}}_{2[23]3}(t_B) \right]_{,[31][13]} - \left[ T_{13}\left( y^{\mathrm{GW}}_{3[31]1}(t_B)\right) + y^{\mathrm{GW}}_{1[13]3}(t_B) \right]_{,[32][23]} \nonumber \\ 
    & - f_{0, [31][13]} + f_{0,[23][32][31][13]} + f_{0, [32][23]} - f_{0,[31][13][32][23]} \nonumber \\ 
    &= T_{23}\left( \left[ 1 + \hat{k}^{\mathrm{GW}}(t_B) \cdot \hat{n}_{[32]}(t_B) \right] \frac{1}{2} \cross \left\{ \Psi_{[32]} \left[ t_3(t_B) - \hat{k}^{\mathrm{GW}}(t_B) \cdot \vec{p}_3(t_3(t_B))\right] - \Psi_{[32]} \left[ t_B - \hat{k}^{\mathrm{GW}}(t_B) \cdot \vec{p}_2(t_B) \right] \right\} \right)_{,[31][13]} \nonumber \\
   &+ \left[ 1 + \hat{k}^{\mathrm{GW}}(t_B) \cdot \hat{n}_{[23]}(t_B) \right] \frac{1}{2} \cross \left\{ \Psi_{[23]} \left[ t_2(t_B) - \hat{k}^{\mathrm{GW}}(t_B) \cdot \vec{p}_2(t_2(t_B))\right] - \Psi_{[23]} \left[ t_B - \hat{k}^{\mathrm{GW}}(t_B) \cdot \vec{p}_3(t_B) \right] \right\}_{,[31][13]} \nonumber \\
   & - T_{13}\left( \left[ 1 + \hat{k}^{\mathrm{GW}}(t_B) \cdot \hat{n}_{[31]}(t_B) \right] \frac{1}{2} \cross \left\{ \Psi_{[31]} \left[ t_3(t_B) - \hat{k_B} \cdot \vec{p}_3(t_3(t_B))\right] - \Psi_{[31]} \left[ t_B - \hat{k}^{\mathrm{GW}}(t_B) \cdot \vec{p}_1(t_B) \right]  \right\} \right)_{,[32][23]} \nonumber \\
   & - \left[ 1 + \hat{k}^{\mathrm{GW}}(t_B) \cdot \hat{n}_{[13]}(t_B) \right] \frac{1}{2} \cross \left\{ \Psi_{[13]} \left[ t_1(t_B) - \hat{k_B} \cdot \vec{p}_1(t_1(t_B))\right] - \Psi_{[13]} \left[ t_B - \hat{k}^{\mathrm{GW}}(t_B) \cdot \vec{p}_3(t_B) \right]  \right\}_{,[32][23]} \nonumber \\
   & -f_{0, [31][13]} + f_{0, [32][23]} - f_{0,[31][13][32][23]} + f_{0,[23][32][31][13]}. 
\end{align}
\end{widetext}
In Eq.~\ref{eq:nswap1c_X_off} above, observe that from the expressions in Eqs.~\ref{eq:n1cH313},~\ref{eq:n1cB323dopplerlink}, we get \textit{eight} terms for the VB signal, whereas the laser phase noise is comparable to the the level as in the classical TDI Michelson Z derived using 6 one-way links. This implies that locking results into \textit{signal enhancement}, ignored by all existing work in this field so far. The signal-to-noise ratio for the choice of the verification binaries is given in Table~\ref{tab:VBSNR}, where comparison with TDI Z derived from free-running lasers is made.
\begin{figure}
  \centering
    \includegraphics[width=0.45\textwidth]{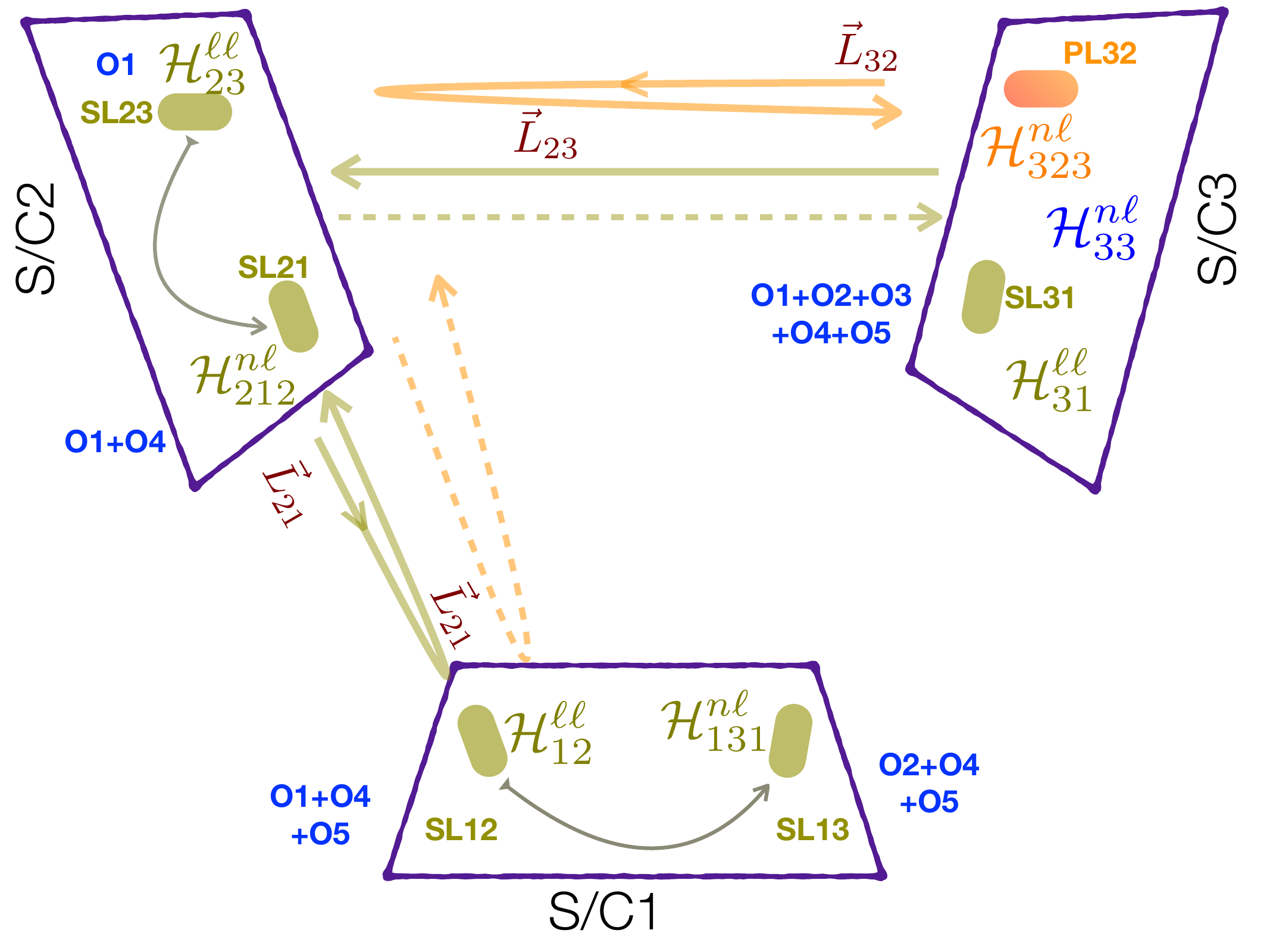}
  \caption{An equivalent `TDI Y' for N2a recovered using phasemeter measurements in solid curves. The delays applied to the measurements are shown in dashed lines.
\label{fig:n2a_Y}}
\end{figure}
\subsection{N2a}
Similarly, we consider which equal pathlength for given measurements for `N2a' result into an equivalent TDI Michelson, preferably the one identical to TDI Z. Inspection of the measurements in Sec.~\ref{n2a_resp} reveals that it is not possible to construct TDI Z centered at S/C 3 since the measurements in that S/C necessarily involve Doppler locks from all three armlengths. However, an observable similar to Z, Michelson ``TDI Y", can be constructed by subtracting $\mathcal{H}^{n\ell}_{323}$ delayed by $T_{21} T_{12}$ from $\mathcal{H}^{n\ell}_{212}$ delayed by $T_{32}$. Along with the long-arm locked and reference measurements to subtract off the offsets we get: 
\begin{widetext}
\begin{align}
    \label{eq:nswap2a_Y}
    Y(t_B)^{\text{N2a}} &=-f_{0, [21][12]} + f_{0,[23][32][21][12]} +f_{0,[32][32]} - f_{0,[12][21][32][32]}  \nonumber \\
    &+D_{32,[23][21][12]} + D_{23, [21][12]} + D_{32, [32]} - D_{32,[12][21][32]} - D_{13,[21][32]} + D_{12, [32]} \nonumber \\
    & + O1_{,[23][21][12]} + O1_{,[32]} - O1_{[12][21][32]} + O4_{,[32]} - O4_{,[12][21][32]} - O5_{,[12][32]}  \nonumber \\
    & + \mathcal{H}^{\ell\ell}_{23, [23][21][12]}+ \mathcal{H}^{\ell\ell}_{23, [32]}- \mathcal{H}^{\ell\ell}_{23, [12][21][32]} - \mathcal{H}^{\ell\ell}_{22, [32]} + \mathcal{H}^{\ell\ell}_{22, [12][21][32]} - \mathcal{H}^{\ell\ell}_{12, [12][32]} \nonumber \\
    &= \mathcal{H}^{n\ell,\text{GW}}_{323, [21][12]} - \mathcal{H}^{n\ell,\text{GW}}_{212, [32]} \nonumber \\
    &-f_{0, [21][12]} + f_{0,[23][32][21][12]} +f_{0,[32][32]} - f_{0,[12][21][32][32]} \nonumber \\
    & T_{23}\left( \left[ 1 + \hat{k}^{\mathrm{GW}}(t_B) \cdot \hat{n}_{[32]}(t_B) \right] \frac{1}{2} \cross \left\{ \Psi_{[32]} \left[ t_3(t_B) - \hat{k}^{\mathrm{GW}}(t_B) \cdot \vec{p}_3(t_3(t_B))\right] - \Psi_{[32]} \left[ t_B - \hat{k}^{\mathrm{GW}}(t_B) \cdot \vec{p}_2(t_B) \right] \right\} \right)_{, [21][12]} \nonumber \\
    &+ \left[ 1 + \hat{k}^{\mathrm{GW}}(t_B) \cdot \hat{n}_{[23]}(t_B) \right] \frac{1}{2} \cross \left\{ \Psi_{[23]} \left[ t_2(t_B) - \hat{k}^{\mathrm{GW}}(t_B) \cdot \vec{p}_2(t_2(t_B))\right] - \Psi_{[23]} \left[ t_B - \hat{k}^{\mathrm{GW}}(t_B) \cdot \vec{p}_3(t_B) \right] \right\}_{, [21][12]} \nonumber \\
    & + \left[ 1 + \hat{k}^{\text{GW}}(t_B) \cdot \hat{n}_{[32]}(t_B) \right]\frac{1}{2} \cross \left\{ \Psi_{[32]} \left[ t_3(t_B) - \hat{k}^{\text{GW}}(t_B) \cdot \vec{p}_3(t_3(t_B))\right] - \Psi_{[32]} \left[ t_B - \hat{k}^{\text{GW}}(t_B) \cdot \vec{p}_2(t_B) \right] \right\}_{, [32]} \nonumber \\
    &- T_{12}\left( T_{21} \left( \left[ 1 + \hat{k}^{\text{GW}}(t_B) \cdot \hat{n}_{[32]}(t_B) \right]\frac{1}{2} \cross \left\{ \Psi_{[32]} \left[ t_3(t_B) - \hat{k}^{\text{GW}}(t_B) \cdot \vec{p}_3(t_3(t_B))\right] - \Psi_{[32]} \left[ t_B - \hat{k}^{\text{GW}}(t_B) \cdot \vec{p}_2(t_B) \right]  \right\} \right) \right)_{, [32]} \nonumber \\
    &- \left[ 1 + \hat{k}^{\text{GW}}(t_B) \cdot \hat{n}_{[21]}(t) \right] \frac{1}{2} \cross \left\{ \Psi_{[21]} \left[ t_2(t_B) - \hat{k}^{\text{GW}}(t_B) \cdot \vec{p}_2(t_2(t_B))\right] - \Psi_{[21]} \left[ t_B - \hat{k}^{\text{GW}}(t_B) \cdot \vec{p}_1(t_B) \right]  \right\}_{, [32]} \nonumber \\
    &- \left[ 1 + \hat{k}^{\text{GW}}(t_B) \cdot \hat{n}_{[12]}(t) \right] \frac{1}{2} \cross \left\{ \Psi_{[12]} \left[ t_1(t_B) - \hat{k}^{\text{GW}}(t_B) \cdot \vec{p}_1(t_1(t_B))\right] - \Psi_{[12]} \left[ t_B - \hat{k} \cdot \vec{p}_2(t_B) \right]  \right\}_{, [32]} \nonumber \\
    &-f_{0, [21][12]} + f_{0,[23][32][21][12]} +f_{0,[32][32]} - f_{0,[12][21][32][32]}.
\end{align}
\end{widetext} 
Similar to the previous sub-section, it is assumed that the residual offset locking noise can be perfectly subtracted using the locking beatnotes, Eqs.~\ref{eq:nswap2a:locrefs}, ~\ref{eq:nswap2a:h23h31h12} for TDI Y. The individual $\Psi$ along the links add up to \textit{twelve} terms. We obtain following signal to noise ratios:
\begin{table}[htpb]
\centering
\begin{tabular}{||c c c||}
\hline
TDI & V407Vul & ZTFJ2243 \\ [0.5ex] 
\hline\hline
free lasers $Z$ & $\approx 37$ &$\approx 38$ \\ 
N1c $Z$ & $\approx 140$ &$\approx 126$ \\ 
N2a $Y$ & $\approx 107$ &$\approx 103$ \\ 
\hline
\end{tabular}
\caption{SNR for GW measurements for the `face-on' V407Vul and the most compact detached know `edge-on' ZTF J1539+5027 binaries from Table~\ref{tab:VBsparam} for observation of 1 year.}
\label{tab:VBSNR}
\end{table}

The SNRs above imply a discovery of significant enhancement of the GW signal caused by laser-locking, using verification binary, which is an ultra-compact binary - not know prior to this work. The enhancement for the two VBs are in the same order, by a factor of 10. Assuming synchronized measurements, and for the fiducial TDI observables, the laser locking configuration with the minimum Doppler links, `N1c', is slightly favourable with higher SNR for both binaries. The SNR difference could be attributable to the different sensitivities of the TDI observables to the GW signals, known previously \cite{Armstrong1999}. A further check is required for noise orthogonal ``A, E, T" type measurements, once they are discovered.
\section{Conclusions} 
\label{conclusion}
A pilot study and preliminary data analysis of the coupling of the Verification Binary signal to the LISA phasemeter data are derived in the case of locking scheme, where all the lasers in the LISA constellation are locked to a primary laser. Many options are available for the locking scheme that can be implemented for any LISA-like geometry. We deviate from the usual assumption of the configuration of the six free-running lasers across the constellation for forming the interferometric measurements with long-lived GW signals in them. We show that contrary to popular belief, the sensitivity of a GW signal is significantly affected by locking scheme, yielding enhanced SNR for the locking configuration compared to that of the free-running lasers. We perform SNR studies for two of the six non-swap locking schemes. These have been implemented in \texttt{Synthetic LISA} \citep{Vallis2005}\footnote{available upon request to lead author}.

From simplified assumptions, we find that at the TDI output, the choice of the minimal locking scheme (N1c) is favoured over the maximal locking  scheme (N2a). Many assumptions have been made such as all the raw measurements have been synchronised \cite{Reinhardt2023} perfectly to $t_B$ and thus this needs further thorough investigation including other laser locking configurations as well. Additionally, we need further analysis with remaining four non-swap configurations to verify the results presented. Furthermore, it is important to stress that verification binaries will provide an \textit{independent} verification of the instrument and calibration of the data's amplitude and phase. 

In phasemeter observation equations, we find that the TDI power for the non-swap locking scheme is significantly larger for edge-on systems and has a significantly smaller dependence on the inclination angle. This has several consequences. First, the known population of verification binaries is biased towards edge-on systems as eclipses are easier to detect in electromagnetic data. Therefore, we expect that most verification binaries will have a significantly larger TDI power for the non-swap locking scheme. Second, the smaller dependence on the inclination angle leads to less bias, based on inclination for newly detected sources. This is particularly important for population studies as they rely on well understood biases in the data. This also means that edge-on Massive Black Hole Binaries (MBHB) would be readily visible in `N2a' raw measurements were the MBHB signals to exceed laser frequency noise at that level, which is perhaps unlikely. But if so, then `N2a' must be in operation for multi-messenger with MBHB type sources.
\section*{Acknowledgement}
SS acknowledges fruitful discussions with Michele Vallisneri for the single link response, Gerhard Heinzel for verifying the \texttt{fplan} lock equations and valuable comments to improve the paper by Kohei Yamamoto, Sarah Paczkowski, and LDPG On-Ground Instrument Processing Expert Group. SS also gratefully acknowledges financial support by the German Aerospace Center (DLR) with funds from the German Federal Ministry for Economic Affairs and Climate Action (BMWK) under grants \# 50 OQ1801 and \# 50 OQ2301. TK acknowledges support from the National Science Foundation through grant AST \#2107982, from NASA through grant 80NSSC22K0338 and from STScI through grant HST-GO-16659.002-A. This result is part of a project that has received funding from the European Research Council (ERC) under the European Union’s Horizon 2020 research and innovation programme (Grant agreement No. 101078773)
\bibliographystyle{apj}
\bibliography{bibl}
\end{document}